\documentclass[structabstract]{aa}  
\usepackage{graphicx}
\usepackage{array}
\usepackage{txfonts}
\usepackage{natbib}
\bibpunct{(}{)}{;}{a}{}{,}

\begin{document}
   \title{Multiwavelength campaign on Mrk 509}
   \subtitle{VIII. Location of the X-ray absorber}

\author{J.S. Kaastra\inst{1,2}
  \and R.G. Detmers\inst{1,2}
  \and M. Mehdipour\inst{3}
  \and N. Arav\inst{4}
  \and E. Behar\inst{5}
  \and S. Bianchi\inst{6}
  \and G. Branduardi-Raymont\inst{3}
  \and M. Cappi\inst{7}
  \and E. Costantini\inst{1}
  \and J. Ebrero\inst{1}
  \and G.A. Kriss\inst{8,9}
  \and S. Paltani\inst{10}
  \and P.-O. Petrucci\inst{11}
  \and C. Pinto\inst{1}
  \and G. Ponti\inst{12}
  \and K.C. Steenbrugge\inst{13,14}
  \and C.P. de Vries\inst{1}
  }
  
\institute{SRON Netherlands Institute for Space Research, Sorbonnelaan 2,
           3584 CA Utrecht, the Netherlands 
	   \and
	   Sterrenkundig Instituut, Universiteit Utrecht, 
	   P.O. Box 80000, 3508 TA Utrecht, the Netherlands
	   \and
	   Mullard Space Science Laboratory, University College London, 
	   Holmbury St. Mary, Dorking, Surrey, RH5 6NT, UK
	   \and
	   Department of Physics, Virginia Tech, Blacksburg, VA 24061, USA
	   \and
	   Department of Physics, Technion-Israel Institute of Technology, 
	   Haifa 32000, Israel 
	   \and 
	   Dipartimento di Fisica, Universit\`a degli Studi Roma Tre, 
	   via della Vasca Navale 84, 00146 Roma, Italy 
           \and
	   INAF-IASF Bologna, Via Gobetti 101, 40129 Bologna, Italy
	   \and
	   Space Telescope Science Institute, 3700 San Martin Drive, 
	   Baltimore, MD 21218, USA
	   \and
	   Department of Physics and Astronomy, The Johns Hopkins University,
	   Baltimore, MD 21218, USA
	   \and
	   ISDC Data Centre for Astrophysics, Astronomical Observatory of the
	   University of Geneva, 16, ch. d'Ecogia, 1290 Versoix, Switzerland 	
	   \and
	   UJF-Grenoble 1 / CNRS-INSU, Institut de Plan\'etologie et d'Astrophysique
	   de Grenoble (IPAG) UMR 5274, Grenoble, F-38041, France
	   \and
	   School of Physics and Astronomy, University of Southampton, 
	   Highfield, Southampton SO17 1BJ, UK
	   \and
           Instituto de Astronom\'ia, Universidad Cat\'olica del Norte, 
	   Avenida Angamos 0610, Casilla 1280, Antofagasta, Chile
	   \and
	   Department of Physics, University of Oxford, Keble Road, 
	   Oxford OX1 3RH, UK
} 
\date{\today}

\abstract
{}
{More than half of all active galactic nuclei show strong photoionised outflows.
We use a massive multiwavelength
monitoring campaign on the bright Seyfert~1 galaxy Mrk~509 to constrain the
location of the outflow components dominating the soft X-ray band.}
{Mrk~509 was monitored by XMM-Newton and other satellites in 2009. We have
studied the response of the photoionised gas to the changes in the ionising flux
produced by the central regions. We used the five discrete ionisation components
A--E that we detected in the time-averaged spectrum taken with the RGS
instrument. By using the ratio of fluxed EPIC-pn and RGS spectra, we were able
to put tight constraints on the variability of the absorbers. Monitoring with
the Swift satellite started six weeks before the XMM-Newton observations. This
allowed us to use the history of the ionising flux and to develop a model for
the time-dependent photoionisation in this source.}
{Components A and B are too weak for variability studies, but the distance for
component A is known from optical imaging of the [\ion{O}{iii}] line to
be about 3~kpc. During the five weeks of the XMM-Newton observations we found no
evidence of changes in the three X-ray dominant ionisation components C, D, and
E, despite a huge soft X-ray intensity increase of 60\% in the middle of our
campaign. This excludes high-density gas close to the black hole. Instead, using
our time-dependent modelling, we find that the density is very low, and we
derive firm lower limits to the distance of these components. For component D we
find evidence for variability on longer time scales by comparing our spectra to
archival data taken in 2000 and 2001, yielding an upper limit to the distance.
For component E we derive an upper limit to the distance based on the argument
that the thickness of the absorbing layer must be less than its distance to the
black hole. Combining these results, at the 90\% confidence level, component C
has a distance of $>70$~pc, component D is between 5--33~pc, and component E has
a distance $>5$~pc but smaller than 21--400~pc, depending upon modelling
details. These results are consistent with the upper limits that we derived from
the HST/COS observations of our campaign and point to an origin of the dominant,
slow ($v<1000$~km\,s$^{-1}$) outflow components in the NLR or torus-region of
Mrk~509.}
{}

\keywords{Galaxies: active --  quasars: absorption lines -- X-rays: general
--- X-rays: galaxies --- Galaxies, individual: Mrk~509}
\maketitle

\section{Introduction}

It has been known for a long time that accretion onto a super-massive black hole
plays a key role in the physics of active galactic nuclei (AGN). While outflows
have already been detected using early optical observations of \object{NGC~4151}
\citep{anderson1969}, their widespread occurrence and role was only recognised
after high-resolution UV spectrographs became available. \citet{crenshaw1999}
found that 60\% of all Seyfert 1 galaxies show UV absorption lines. While
intrinsic X-ray absorption had already been found in the early 1980s, including
the first evidence of absorption from ionised gas \citep{halpern1984}, the proof
that this gas is outflowing came from the first high-resolution X-ray spectrum
of a Seyfert galaxy \citep{kaastra2000}. We refer to \citet{crenshaw2003} for a
more extensive overview of outflows.

Measurements of absorption line spectra yield reliable information on different
aspects of the outflow such as dynamics (through the Doppler shifts and line
profiles) and ionisation state (through intensity ratios of lines from different
ions). However, they do not provide a key ingredient, the distance $r$ of the
absorbing material to the central black hole. The basic problem is that the
ionisation state is described by the parameter $\xi=L/nr^2$ with $L$ the
ionising luminosity between 1--1000~Ryd and $n$ the hydrogen density. With $L$
and $\xi$ known from observations, only the product $nr^2$ is known. This
degeneracy has led to considering many different locations for the outflow,
including the halo of the host galaxy (10~kpc), the inner galactic disc (0.1--1
kpc), the narrow emission line region (10--500~pc), and the intermediate line
region (0.01--10~pc). The recently discussed, so-called ultra-fast outflows
\citep[see e.g.][and references therein]{tombesi2010} may be even further
inwards. 

There are several methods of estimating $r$. The most direct and easiest method
is to use direct imaging. A good example are the Doppler maps of [\ion{O}{iii}]
$\lambda5007$ by \citet{phillips1983} in Mrk~509. However, this method can only
be applied for the most nearby sources and for the most extended low-ionisation
gas. Instrumental capabilities are insufficient to resolve more centrally
concentrated gas.

Other methods rely on obtaining the density $n$ and deriving the distance $r$
through using the ionisation parameter. The most direct way is to use
density-sensitive lines from meta-stable levels. This method has been applied in
a few cases using different UV lines from meta-stable levels in various sources.
The published results display a broad range of distances for different sources,
from $<$0.1~pc \citep{kraemer2006}, 25~pc \citep{gabel2005} to 3~kpc
\citep{moe2009}. In X-rays, \citet{kaastra2004} pioneered this method
theoretically using K-shell absorption lines from \ion{O}{v}, but owing to the
limited signal from the target that was studied, \object{Mrk~279}, they could
not obtain firm detections of these lines. For X-ray binaries, however,
metastable lines have been found in \object{GRO~J1655--40} \citep{miller2008},
but they sample a density range ($\sim$$10^{20}$~m$^{-3}$) that is too high for
AGN.

Alternatively, variations in the ionising luminosity of an AGN will induce
changes in the ionic composition of the surrounding gas, hence changes in the
transmitted spectrum. How fast the gas responds depends on the recombination
time scale, which is a function of density (for absorption the light-crossing
time is not relevant). By measuring the delay (or lack of delay) of an
absorption component relative to the luminosity variations, this recombination
time scale can be determined and from that the density $n$ and distance $r$ are
deduced. The method has been applied to several sources, again yielding a broad
range of distances \citep[see for references our introduction in paper I of this
series,][]{kaastra2011a}.

In this paper we follow the last method. In 2009 we have performed a large
multiwavelength campaign on one of the best targets for this study,
\object{Mrk~509} \citep{kaastra2011a}. The source has a typical variability time
scale of a few days, with sufficiently large amplitude. The first step of our
analysis was to characterise the time-averaged properties of the absorber using
the combined RGS spectrum of all observations taken by XMM-Newton during our
campaign. This has been described in detail by \citet{detmers2011}. Using the
multiwavelength monitoring as described by \citet{kaastra2011a}, it is then
possible to predict the variations of the transmission of the outflow for a grid
of possible distances $r$. Comparing these predictions with the observed spectra
then yields distances to the outflow components. In our comparison, we have
focussed on the data obtained with the pn camera of XMM-Newton, because that has
the highest sensitivity to variations in the 1~keV band. The MOS data could not
be used for this comparison because of a small amount of pile-up. However, we
also obtained useful constraints from the RGS spectra at longer wavelengths.

The structure of this paper is as follows. First we summarise the ionisation
structure of the outflow, investigate the spectral signatures of the expected
variations of the transmission, and introduce the data sets that we use
(Sect.~\ref{sect:data}). As a next step (Sect.~\ref{sect:denslim}) we compare
the data to the predictions for the high-density limit (instantaneous response
of the absorber to continuum variations). Afterwards we describe our method of
treating the more general case of time-dependent photoionisation
(Sect.~\ref{sect:timedependent}). In Sect.~\ref{sect:longterm} we compare the
data from our 2009 campaign to archival data to obtain constraints from the
long-term variability. We discuss our results briefly in
Sect.~\ref{sect:discussion}.

We use 1$\sigma$ (68\% confidence) uncertainties throughout this paper, unless
stated explicitly otherwise.

\section{Data\label{sect:data}}

\subsection{Fluxed spectra}

We have made fluxed spectra for each of the ten observations as well as for the
combined spectrum for each of the instruments available to us. The need to use
fluxed spectra is explained in more detail in Sect.~\ref{sect:method} and is
related to the weakness of the expected signals. More details about the 
observations and the observation log are described by \citet{kaastra2011a}.

\subsubsection{EPIC pn spectra\label{sect:epicpn}}

The pn \citep{strueder2001} count spectra were extracted as described by Ponti
et al. (2011, in preparation). We have produced fluxed spectra from these count
rate spectra using the program {\sl epicfluxer} that is available in the public
SPEX\footnote{www.sron.nl/spex} distribution. Our procedure is as follows. We
rebin the count spectra energy-dependent to about 1/3 FWHM, ignore the parts
below 0.2~keV and above 12~keV, and fit the count spectra using a cubic spline
(the {\sl spln} model of SPEX) on a logarithmic energy grid between 0.2--12~keV
with 100 grid points. The resolution of this grid is about 2\%, and it matches
the spectral resolution of pn well. All fits are statistically acceptable. We
then take the corresponding model photon spectra (also as splines) and convolve
them with a Gaussian redistribution function. The FWHM of this Gaussian is
energy-dependent and corresponds to the spectral resolution of the instrument at
each energy. In this way we have a model spectrum at EPIC resolution that
describes the data perfectly. Any remaining residuals that are visible in a plot
of the original fit to the count spectra with the \textsl{spln} model, come
mostly from the statistical noise of the data. To take that noise into account
and also to correct for any remaining weak residuals caused by the limitations
of the \textsl{spline} photon spectrum approximation to the true photon
spectrum, we scale the convolved spline model spectra with the ratio of the
observed data to the predicted data at each energy.

\begin{figure}[!tbp]
\resizebox{\hsize}{!}{\includegraphics[angle=-90]{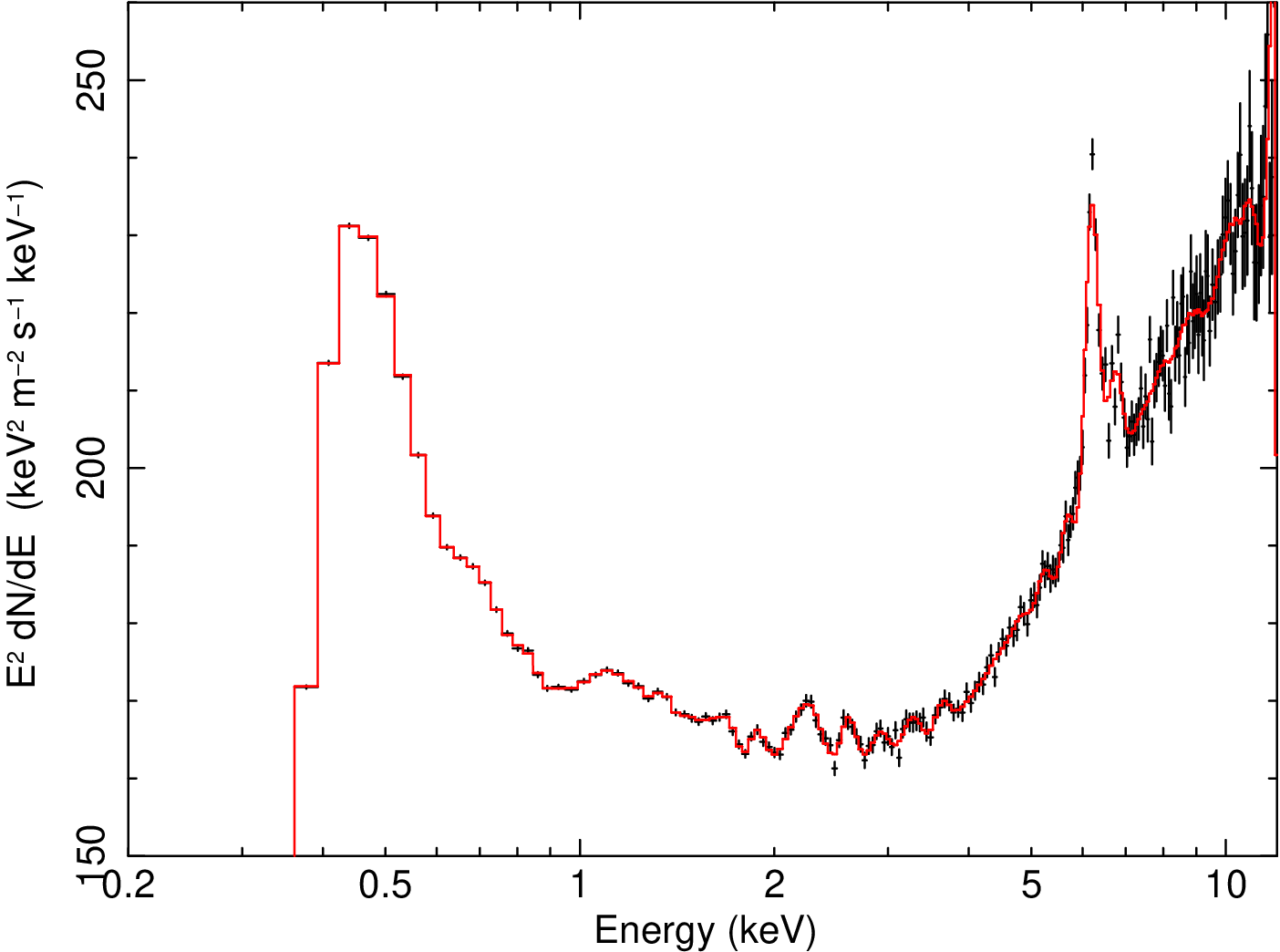}}
\caption{Fluxed pn spectrum for the co-added 2009 spectrum of Mrk~509. The
spectrum has been multiplied by $E^2$ for clarity in the display.}
\label{fig:efluxepn}
\end{figure}

Fig.~\ref{fig:efluxepn} shows an example of the fluxed spectrum of the combined
2009 data. For our analysis we used the data points, but the plot shows the
excellent agreement between the response-convolved spline model and the data.
The spectrum clearly shows the Fe-K line near 6.4~keV, as well as a dip caused
by the absorbing outflow between 0.8 and 1.0 keV. Some of the features between
1.8 and 4 keV may be caused by remaining small calibration uncertainties at a
level of a few percent, and are possibly associated to the gold edge of the
mirror or the silicon edge of the detector. However, whether these features are
instrumental or have an astrophysical origin is not important for the present
purpose, as  they disappear in the ratio of different spectra. Below 0.4~keV the
flux drops rapidly owing to the Galactic absorption. Since the spectral energy
resolution in that region of the spectrum is low anyhow and we do not expect to
see significant variations in the absorption components in that band, it is not
important for the present paper, so we ignore the spectrum below 0.5~keV in the
rest of the paper.

\begin{figure}[!tbp]
\resizebox{\hsize}{!}{\includegraphics[angle=-90]{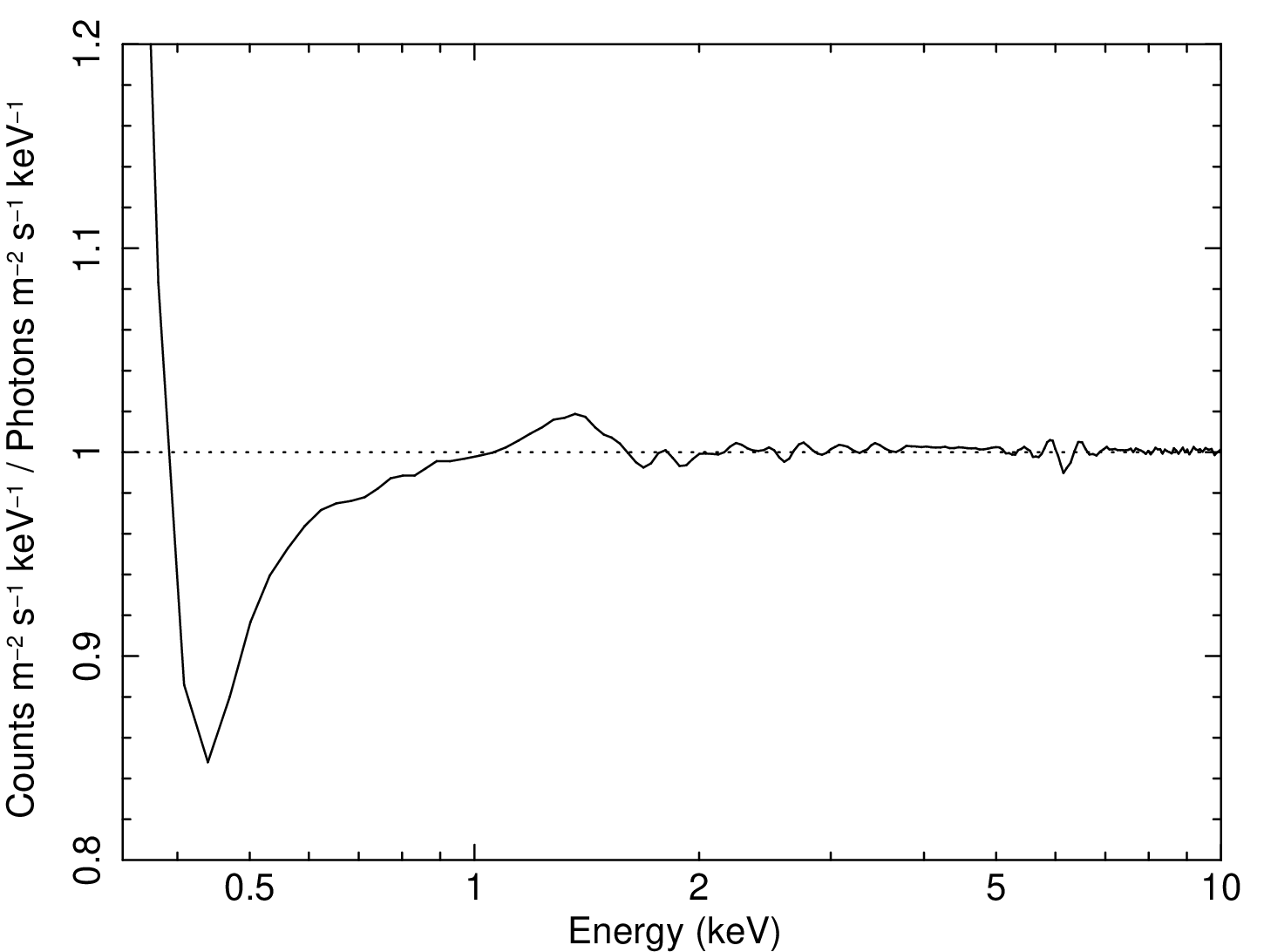}}
\caption{Ratio of the `fluxed' pn count spectrum to the fluxed photon spectrum
of Mrk~509.}
\label{fig:zplot}
\end{figure}

Furthermore, we take the following into consideration. Fig.~\ref{fig:zplot}
shows the ratio of the nominal count spectrum (counts\,s$^{-1}$\,keV$^{-1}$)
divided by the nominal effective area at each energy, to the fluxed spectrum
produced by the \textsl{epicfluxer} program as described above. In particular,
at lower energies most of the counts correspond to redistributed photons of
higher energies. Because the intrinsic photon spectrum becomes weak at low
energies owing to interstellar absorption, the strong rise of the ratio below
0.35~keV, up to $\sim 4$ for $E=0.2$~keV, is produced. The negative dip around
0.45~keV is caused by loss of photons of 0.45~keV to counts at energies below
0.45~keV.

In the data, an excess of 1\% above the model at 0.5 keV would actually be an
excess of 1\% counts. With the ratio of $\sim 0.9$ at 0.5~keV from
Fig.~\ref{fig:zplot}, this corresponds to an excess in flux of about 1.1\%. Our
procedure would thus give a systematic bias of 0.1\% for the above example. 
However, the typical statistical uncertainty of the spectra in this region is
much smaller than the 1\% of the example ($<$0.4--0.5\%). Furthermore, the
statistical deviations for neighbouring bins are independent and can have
different signs, so their average value over a certain band of bins is even
lower. We conclude that these effects on our analysis can be ignored safely.

The pn is known to be extremely stable over time. Here we quantify this.
Stability can only be measured using unvarying sources. Such sources are
relatively rare, but perhaps the best candidate is the isolated neutron star
\object{RXJ~1856.6--3754}. This object is being used as a regular calibration
target for XMM-Newton. The best band for checking stability is the 0.3--0.5~keV
band, given the soft spectrum of the source. We use the products that are
available with the \textsl{calibration review tool} on the XMM-Newton web page.
The 15 pn observations of this source in the 0.3--0.5~keV band have fluxes with
a standard deviation of 3.8\% around the mean value. We are more interested in
relative differences rather than absolute fluxes, so we have also determined the
mean and standard deviation for the fit residuals averaged over a 0.05~keV wide
band around 0.3~keV. These residuals have a mean value of 2.4\% (consistent with
the nominal calibration uncertainty), but more importantly the scatter (standard
deviation) for individual observations around this mean is only $\sim 0.5\%$.
The true number may be even smaller because this number agrees fully with the
uncertainty due to the photon statistics in this band (it is limited by the
exposure times of the calibration observations). The number implies that
\textsl{relative} time variations in narrow bands can be measured down to the
$\sim 0.5$\% level or better, provided the source has enough photons.

\subsubsection{EPIC MOS spectra}

The MOS \citep{turner2001} data cannot be used for our purpose. In contrast to
the pn data, the ratios between MOS spectra often show weak breaks around
0.7~keV, with deviations below that energy from the extrapolated power law from
the 0.7--1.5 keV band and also deviations from the ratio obtained from the
simultaneous pn data. This is caused by a combination of small amounts of
pile-up, small imperfections in the MOS redistribution function at low energies,
and uncertainties near the instrumental oxygen edge. This softening or hardening
near 0.7~keV makes detection of the expected narrow Gaussian differences in
transmission harder, because at least for the ionisation components C and D, the
centroids of the Gaussians are near this energy (0.716 and 0.820~keV,
Table~\ref{tab:gausblend}). The signal-to-noise ratio of the combined MOS data
in the relevant band is also about two times lower than for the corresponding pn
data. This is partly due to the smaller extraction radii for the MOS spectra
(20\arcsec) as compared to the pn spectra (45\arcsec). For these reasons we do
not include the MOS data in our subsequent analysis.

\subsubsection{RGS spectra}

The data reduction of the RGS \citep{denherder2001} spectra was described in
detail by \citet{kaastra2011b}. The spectra were taken in the so-called
multi-pointing mode. For each observation, one of five offset pointings in the
dispersion direction was chosen. This has the advantages that (i) the effects of
spectral bins with lower exposure from the excision of bad pixels is strongly
reduced in the stacked RGS spectrum, and (ii) that the gaps between CCDs are
filled with data. For individual spectra, this advantage is lost, although it is
partly recovered by the combination of RGS1 and RGS2 data and first- and
second-order spectra.

\begin{figure}[!tbp]
\resizebox{\hsize}{!}{\includegraphics[angle=-90]{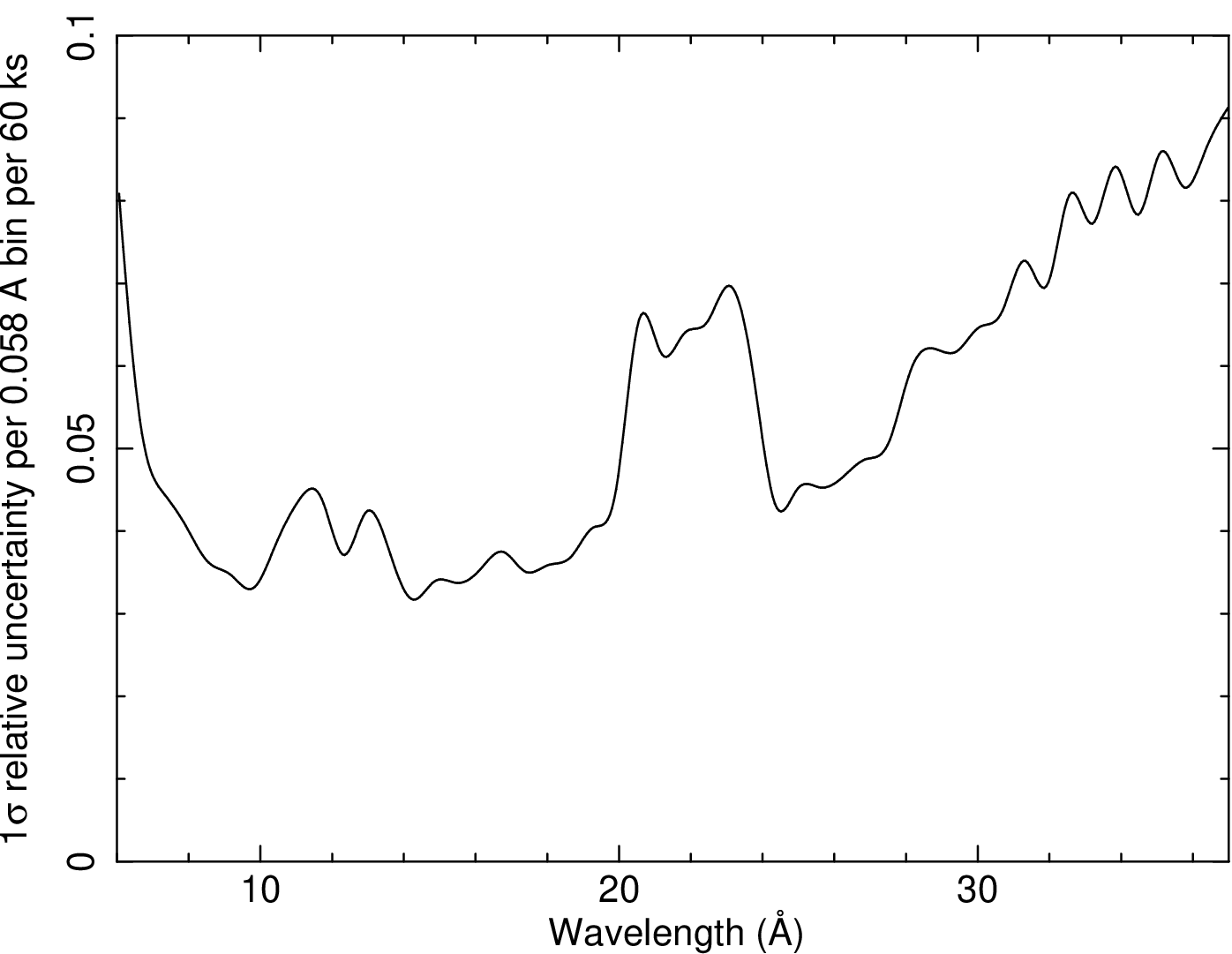}}
\caption{1$\sigma$ relative uncertainty on the individual $\sim$60~ks RGS
spectra of Mrk~509 per 0.058~\AA\ bin.}
\label{fig:rgssens}
\end{figure}

We are searching for variation in the depth of absorption lines, so we have
rebinned our spectra by a factor of 6 to 0.058~\AA, approximately the resolution
(FWHM) of the RGS. In Fig.~\ref{fig:rgssens} we show our sensitivity to
detecting changes in individual RGS spectra. The curve is derived from a
50-point spline fit to the median statistical uncertainties of the ten
individual spectra. It  ranges between 3 and 6\% in the 10--30~\AA\ band, with
some maxima (7--9\%) at the two ends of the band and in correspondence with the
missing CCDs.

\section{Description of the outflow and analysis method\label{sect:method}}

\subsection{Ionisation structure of the outflow}

The time-averaged RGS spectrum of Mrk~509 as analysed by \citet{detmers2011}
shows a number of discrete ionisation components, labelled A--E (see
Table~\ref{tab:components}, taken from the absorption measure distribution fit).
The discrete nature of these components could be proved -- at least for
components C and D -- by applying a continuous model for the column density
versus ionisation parameter $\xi$. That continuous model shows two strong peaks
in absorption measure at the position of components C and D. The statistics of
the spectrum and the lack of suitable strong lines did not allow us to prove the
discrete nature of components A, B, and E, but there is also no strong evidence
against the discrete nature of these components, so for our present analysis we
started with the five components given by \citet{detmers2011}. We use the
parameters from his Table~6, given here in Table~\ref{tab:components}, which is
based on a fit of the measured ionic column densities co-added over all
velocities resulting in five ionisation components. 

\begin{table}[!tbp]
\caption{Ionisation components used in the present analysis} 
\centering                
\begin{tabular}{ccc} 
\hline\hline
component & $\log\xi$\tablefootmark{a} & $N_{\rm H}$\tablefootmark{b} \\
\hline
A & $-0.33\pm 0.49$ & $0.23\pm 0.09$ \\
B & $0.71\pm 0.12$ & $0.84\pm 0.10$ \\
C & $2.01\pm 0.02$ & $4.8\pm 0.4$ \\
D & $2.79\pm 0.06$ & $5.7\pm 0.9$ \\
E & $3.60\pm 0.27$ & $54\pm 73$ \\
\hline
\label{tab:components}                         
\end{tabular}
\tablefoot{
\tablefoottext{a}{Ionisation parameter in $10^{-9}$~W\,m as used throughout
this paper.}
\tablefoottext{b}{Total hydrogen column density in $10^{24}$~m$^{-2}$.}
}\end{table}   
        
Alternatively, \citet{detmers2011} also fitted the spectrum directly with the
\textsl{xabs} model of SPEX \citep{kaastra1996} using multiple discrete
ionisation and velocity components. In that analysis, component C is split over
two velocities ($-40$ and $-270$~km\,s$^{-1}$), with slightly different
$\xi$-values \citep[see Table~5 of][]{detmers2011}. In our discussion we come
back to this issue in the context of discussing the uniqueness of our results.

\subsection{Transmission of the five components\label{sect:trans}}

\begin{figure}[!tbp]
\resizebox{\hsize}{!}{\includegraphics[angle=-90]{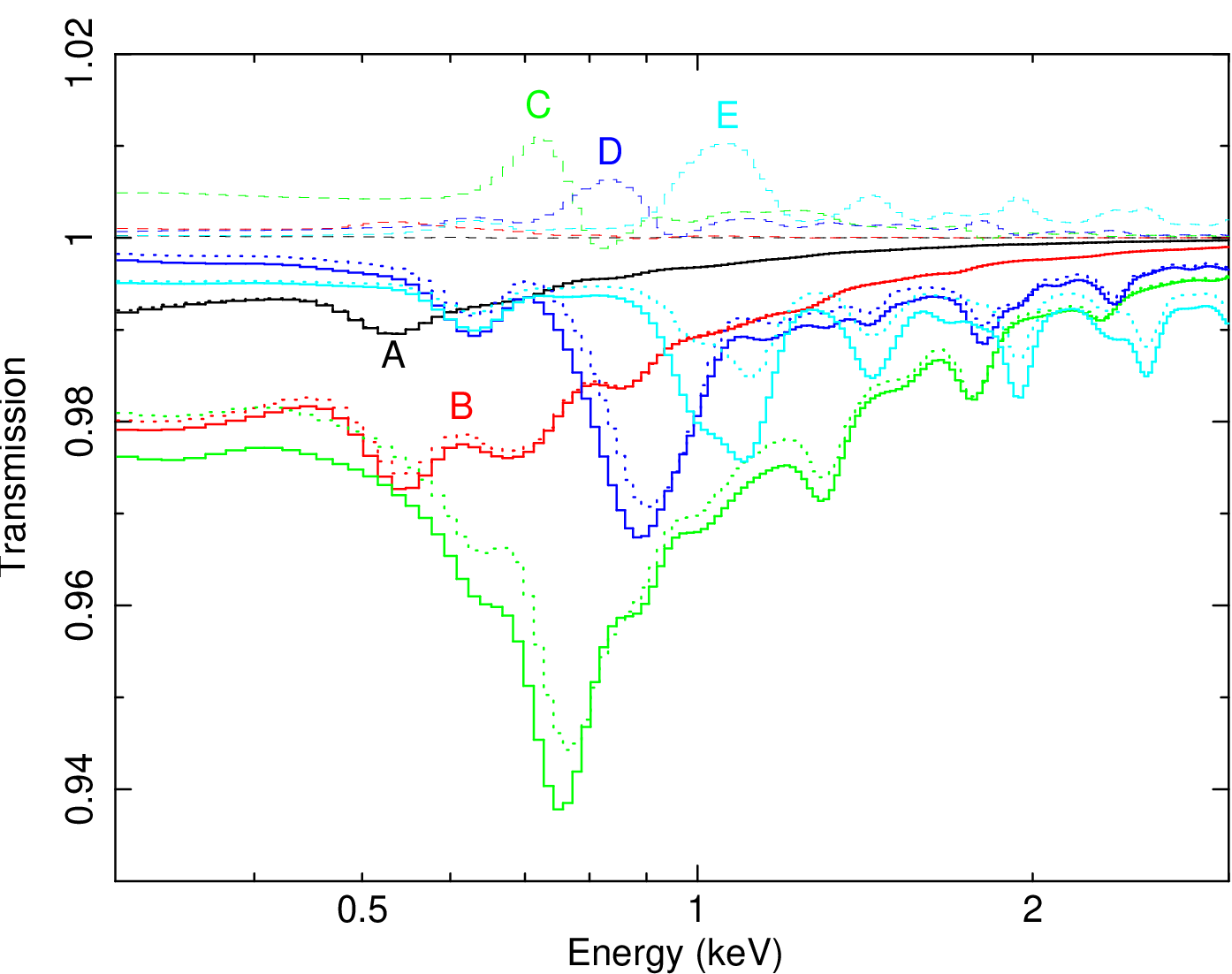}}
\caption{Transmission for the five ionisation components A--E at the 
resolution of the pn instrument (solid lines). Dotted lines: the same but
for $\log\xi$ 0.1 dex higher. Dashed lines: ratios of the transmission
for the higher of the above ionisation states to the lower ionisation states.
We use the same colour coding for each component.}
\label{fig:transchange}
\end{figure}

As a starting point in our analysis, we calculated the transmission, convolved
with the resolution of the pn detector and with the cosmological redshift
applied, for all five individual ionisation components. We focus here on the pn
data because of its higher sensitivity at higher energies, but later we also
consider the RGS spectra. Because Mrk~509 shows variations of 20--25\% in
luminosity ($\sim$0.1~dex),  we also calculated the transmission for these
components using the same total hydrogen column density but with $\log\xi$
0.1~dex higher (recall $\xi=L/nr^2$). For other, not too large variability
amplitudes we can easily scale these calculations. The results are shown in
Fig.~\ref{fig:transchange}. This figure also shows the relative change for each
component if the ionisation balance is adjusted, by plotting the ratio between
the transmission for both cases. 

The figure shows that -- at the pn resolution -- all components show deviations
from unity in the transmission, up to 6\% for component C. However, the
differences in the transmission between the two cases are smaller than the
deviations from unity, and they do not always peak at the energy of the dominant
absorption features. For instance, component D has the strongest absorption
blend at an energy of 0.90 keV, but by increasing the ionisation parameter the
main effect is to reduce the depth of the low-energy wing of the blend,
resulting in a peak of the difference spectrum at 0.82 keV. 

In general ions that have their peak concentration exactly at the ionisation
parameter of one of the detected ionisation components, thereby making the
strongest contribution to the absorption, are not very sensitive to modest
changes in the ionisation parameter, since the second derivative of the ionic
column density versus ionisation parameter is close to zero. Therefore, changes
in the individual components can be spotted most easily using ions that peak
near the relevant ionisation parameter (to yield significant column densities
and absorption features), but not exactly at this ionisation parameter (and
therefore have a strong gradient of the derivative of the column density versus
$\xi$).

\begin{table}[!tbp]
\caption{Approximation with Gaussian components of the changes in transmission 
due to an increase of 0.1 dex in $\log\xi$.} 
\centering                        
\begin{tabular}{lccc} 
\hline\hline
Component & $E_0$ & $\sigma$ & $N$ \\
& (keV) & (keV) & \\
\hline
C & 0.716 & 0.041 & 0.0113 \\
D & 0.820 & 0.073 & 0.0060 \\
E & 1.056 & 0.098 & 0.0102 \\
\hline 
\label{tab:gausblend}                         
\end{tabular}
\end{table}

We found that the prominent peaks due to a change $\Delta T$ in the transmission
for components C, D, and E are represented well by Gaussians:
\begin{equation}
\Delta T = N {\rm e}^{\displaystyle{-(E-E_0)^2/2\sigma^2}}.
\end{equation}
The parameters for these Gaussians (for an increase of 0.1 in $\log\xi$)
are given in Table~\ref{tab:gausblend}.

We discuss the different components in more detail below. Although they affect
the transmission below 1~keV at the 1--2\% level, components A and B are
expected to show changes in the transmission (for an increase in $\log\xi$ by
0.1) no larger than 0.1\%, so cannot be used for the present purpose of
constraining the spectral variations of the absorption components. This lack of
sensitivity is due to the lower column densities for these low-ionisation
components, combined with the poor spectral resolution of EPIC at low energies,
causing relatively strong broadening of spectral lines.

\begin{table}[!tbp]
\caption{Strongest contributions to the dominant absorption troughs for
components C, D, and E.} 
\centering                
\begin{tabular}{@{\,}llccccc@{\,}} 
\hline\hline
Comp & ion \& dominant    & $\lambda$\tablefootmark{a} & $E$\tablefootmark{b}
 & $W_1$\tablefootmark{c} &  $W_2$\tablefootmark{d} & 
$\Delta W$\tablefootmark{e} \\
 & transition / blend & (\AA) & (keV) & (eV) & (eV) & (eV) \\
\hline
C & \ion{O}{vii} 1s--3p$\ldots$7p& 18.63 & 0.663 & 0.99 & 0.75 & $-0.24$ \\
C & \ion{Fe}{ix}                 & 16.54 & 0.721 & 0.17 & 0.06 & $-0.11$ \\
C & \ion{Fe}{x}                  & 16.36 & 0.729 & 0.52 & 0.23 & $-0.29$ \\
C & \ion{Fe}{xi}                 & 16.18 & 0.738 & 0.84 & 0.43 & $-0.41$ \\
C & \ion{Fe}{xii}                & 16.03 & 0.745 & 0.69 & 0.53 & $-0.16$ \\
C & \ion{Fe}{xiii}               & 15.89 & 0.756 & 0.60 & 0.55 & $-0.05$ \\
C & \ion{Fe}{xiv}                & 15.63 & 0.767 & 0.27 & 0.32 & $+0.05$ \\
C & \ion{O}{viii} 1s--3p,4p,5p   & 16.01 & 0.768 & 1.58 & 1.50 & $-0.06$ \\
C & \ion{Fe}{xv}                 & 15.36 & 0.777 & 0.16 & 0.23 & $+0.07$ \\
C & \ion{Fe}{xvi}                & 15.26 & 0.786 & 0.13 & 0.22 & $+0.09$ \\
C & \ion{Fe}{xvii} 2p--3s,3d     & 15.01 & 0.789 & 0.57 & 1.07 & $+0.50$ \\
\hline
D & \ion{O}{viii} 1s--3p,4p,5p&16.01& 0.766 & 0.41 & 0.26 & $-0.15$ \\
D & \ion{Fe}{xvii}          & 15.01 & 0.794 & 0.47 & 0.25 & $-0.22$ \\
D & \ion{Fe}{xviii}         & 14.20 & 0.836 & 1.27 & 0.89 & $-0.38$ \\
D & \ion{Fe}{xix}           & 13.52 & 0.881 & 1.54 & 1.41 & $-0.13$ \\
D & \ion{Ne}{ix}    1s--2p  & 13.45 & 0.891 & 0.20 & 0.11 & $-0.09$ \\
D & \ion{Fe}{xx}    2p-3d   & 12.84 & 0.929 & 0.65 & 0.86 & $+0.21$ \\
\hline
E & \ion{Fe}{xx}    2p--3d  & 12.84 & 0.935 & 0.07 & 0.00 & $-0.07$ \\
E & \ion{Ne}{x}     1s--2p  & 12.13 & 0.987 & 0.66 & 0.46 & $-0.16$ \\
E & \ion{Fe}{xxi}   2p--3d  & 12.29 & 0.987 & 0.40 & 0.07 & $-0.33$ \\
E & \ion{Fe}{xxii}  2s--3p  & 11.71 & 1.038 & 0.84 & 0.19 & $-0.65$ \\
E & \ion{Fe}{xxiii} 2s--3p  & 10.98 & 1.092 & 1.01 & 0.47 & $-0.54$ \\
E & \ion{Fe}{xxiv}  2s--3p  & 10.62 & 1.127 & 1.06 & 0.76 & $-0.30$ \\
\hline
\label{tab:line}                         
\end{tabular}
\tablefoot{
\tablefoottext{a}{Laboratory wavelength of the strongest 
individual spectral line of the ion that contributes to the blend.}
\tablefoottext{b}{Predicted energy of the line centroid of all lines from the given ion
that contribute to the blend, in the observer's frame (thus accounting for
cosmological redshift).}
\tablefoottext{c}{Total equivalent width of all the lines of the 
ion contributing to the blend, for the parameters of the time-averaged 
spectrum (see Table~\ref{tab:components}).}
\tablefoottext{d}{As above, but for $\log\xi$ higher by 0.1 dex.}
\tablefoottext{e}{Difference in equivalent width for the case with 
a higher ionisation parameter compared with the time-averaged spectrum.}
}\end{table}                        

For this reason we concentrated on components C, D, and E. In their expected
response to changes in the ionisation parameter, they show significant features
with amplitudes of about a percent at energies of 0.716, 0.820, and 1.056~keV,
respectively. Using our spectral model, we estimated the equivalent widths of
the lines from all ions that contribute to these features. The results are shown
in Table~\ref{tab:line}.

The expected changes in {\sl component C} are dominated by the declining column
densities of \ion{Fe}{ix}--\ion{Fe}{xii} as a function of ionisation parameter,
and also by the Rydberg series of \ion{O}{vii} starting from $n=3$. The 1s--2p
line of this ion is well separated from the presently considered blend. At the
far blue side of the profile, the increase in \ion{Fe}{xvii} column density
enhances the contrast for these column densities of \ion{Fe}{ix}--\ion{Fe}{xii}
slightly. The dominant lines from \ion{Fe}{xv}--\ion{Fe}{xvii} are separated
reasonably well in energy from the more lowly ionised ions, and thus do not
contribute much to the central regions of the blend. As a result, mainly the
changes in \ion{Fe}{ix}--\ion{Fe}{xiv} and \ion{O}{vii} and \ion{O}{viii} for
this component are relevant.

Below 0.715~keV a reduction in the optical depth $\tau$ of the \ion{O}{vii} edge
is also visible (Fig.~\ref{fig:transchange}). It decreases from 0.0112 to
0.0079, causing the transmission above the edge to increase by 0.33\% relative
to the lower energies.

For {\sl component D}, the \ion{O}{viii} 1s--2p line, although it is the
strongest line in this energy band, does not contribute significantly to the
main component of the variation, because it is 150 eV off the centroid and
because the optical depth is large (2.2), causing moderate saturation effects,
hence less sensitivity to column density variations. We have included
\ion{Fe}{xx} for component D in Table~\ref{tab:line}, because this ion is
expected to increase its column density as a response to an increase in
ionisation parameter, resulting in a shift  of the centroid of the varying
transmission component to slightly lower energies.

The changes in {\sl component E} are dominated by decreasing column densities of
\ion{Fe}{xxi}--\ion{Fe}{xxiv} as a function of ionisation parameter. Component E
is also predicted to have a strong absorption feature in the Fe-K band, mainly
due to \ion{Fe}{xxv}. At EPIC resolution, this line has a maximal depth of about
5~\%. Because it is possibly blended by emission features of the same ion, it is
not possible to measure its equivalent width. If component E responds to the
continuum changes, then the line's peak intensity will change by about 1~\%,
which is too small to detect since the signal-to-noise ratio per observation and
per resolution element (FWHM) is only 50 at those high energies. Therefore we do
not consider the Fe-K band further here.

For component E, there is a significant uncertainty of $0.27$ in the value of
$\log\xi$ for the time-averaged spectrum as determined by \citet{detmers2011}.
This is because of a strong correlation between the total hydrogen column
density and $\xi$. The consequences of this on our analysis are discussed in
more detail in Sect.~\ref{sect:discussion}. For components C and D, the
uncertainties are much smaller. However, because we use in this paper changes
\textsl{relative} to the time-averaged spectrum, uncertainties in the absolute
values of $\xi$ and $N_{\rm H}$ for the time-averaged spectrum are less
important.

\subsection{Measuring transmission changes of the outflow}

The continuum flux during the XMM-Newton campaign of Mrk~509 varied
significantly at a level of the order of 25\%, corresponding to 0.1 dex, in the
hard X-ray band \citep{kaastra2011a}. In the soft X-ray band we even see
enhancements up to $\sim$60\% (see Fig.~\ref{fig:lumplot} later in this paper).
If any of the ionisation components of the outflow responds rapidly to the
changing ionising continuum, then we expect to see the changes in the spectrum
as displayed in Fig.~\ref{fig:transchange}. If on the other hand the response
time is slow, no changes in the transmission may occur.
Fig.~\ref{fig:transchange} shows that the expected peaks for components C, D,
and E are well separated in energy, implying that the (lack of) variations can
be detected for each of these components individually.

The amplitude of the expected signal is of the order of 1\%. This is smaller
than the typical systematic uncertainty in the EPIC effective area, which is a
few percent or even slightly larger, and varies with energy. As a result small
differences in the model photon spectra cannot be distinguished from
imperfections in the adopted effective area. Therefore, global spectral fitting
of individual spectra at this level of accuracy is not possible. However, the pn
detector of XMM-Newton is extremely stable (Sect.~\ref{sect:epicpn}), and
therefore \textsl{relative} changes in the spectrum are much easier to detect.

We proceed as follows. We produce fluxed spectra $S_i$ for all ten individual
observations ($i=1,\ldots 10$) as described in the next subsection. The ratios
$r_{ij,{\rm{obs}}}=S_i/S_j$ of these fluxed spectra are then independent of any
effective area uncertainties. Also, in this way the effects of interstellar
absorption are taken out. The expected ratio is
$r_{ij,{\rm{pred}}}=C_iT_i/C_jT_j$, where $C_i$ is the intrinsic continuum for
spectrum $i$ and $T_i$ the transmission of the absorbing component(s). If there
is no change of the absorber, $r_{ij,{\rm{pred}}}$ is just the ratio of the
model continuum spectra. In order to take out the time variability of the
continuum, we use instead
\begin{equation}
R_{ij,{\rm{obs}}}=(S_i/S_j) / (C_i/C_j) - 1,
\label{eqn:robs}
\end{equation}
and compare this with
\begin{equation}
R_{ij,{\rm{pred}}}=T_i/T_j - 1.
\label{eqn:rpred}
\end{equation}
If any of the absorbing components varies over time, $R_{ij,{\rm{pred}}}$ and
the  corresponding $R_{ij,{\rm{obs}}}$ will differ from zero in narrow energy
bands, cf. our findings in Sect.~\ref{sect:trans}.

\section{Searching for variations in the outflow in the high density
limit\label{sect:denslim}}

In the previous section we described the structure of the outflow, indicated the
characteristic predicted variations as a response to changes in the ionising
radiation, and presented the data products that we use to measure these
variations.

In our analysis we use three levels of sophistication. At the most simple level
we assume that the ionising continuum varies only in flux, but not in shape. In
the case of high density, this causes an immediate response to the flux
variations with the characteristic signatures as presented in
Fig.~\ref{fig:transchange}. For gas at a fixed density and distance to the
ionising source, a change of $L$ is then directly proportional to a change in
$\xi$. 

\begin{figure}[!tbp]
\resizebox{\hsize}{!}{\includegraphics[angle=-90]{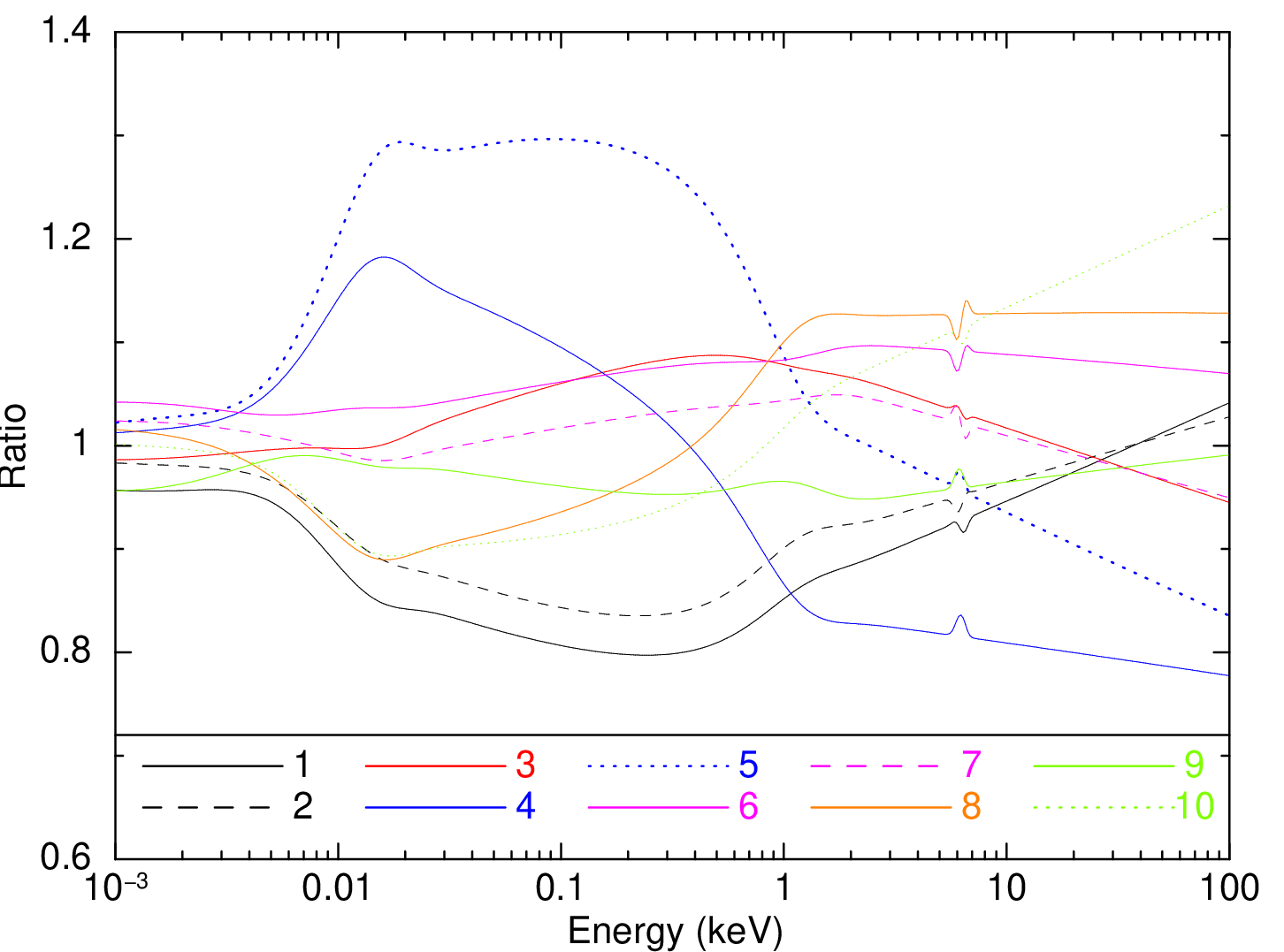}}
\caption{Ratio of the ten model spectra to the time-averaged model
spectrum, based on the fits by \citet{mehdipour2011}. Only the underlying
continua are taken into account, not the absorption components. The spectra
as displayed are extrapolated outside the XMM-Newton range for illustrative
purposes.}
\label{fig:modelrat}
\end{figure}

However, the above situation is not realistic, \citet{mehdipour2011} have shown
that there are significant changes of the spectral shape during our campaign
(Fig.~\ref{fig:modelrat}, Sect.~\ref{sect:ionisingcontinuum}). Therefore in this
section we study a model where the gas has a high density and hence responds
immediately to variations in the ionising flux, but we take the variations in
spectral shape as measured simultaneously with the EPIC and OM instruments of
XMM-Newton into account. This is an important limit for the more general model
of time-dependent ionisation at lower densities that we discuss in
Sect.~\ref{sect:timedependent}.

\subsection{Ionising continuum\label{sect:ionisingcontinuum}}

We follow \citet{mehdipour2011}, who model the continuum as the sum of four
components: a disc blackbody (\textsl{dbb}), a soft excess parametrised by a
Comptonised disc component (\textsl{se}), a power law (\textsl{pl}) and an Fe-K
line (\textsl{fe}, see their Table~5). The iron line does not contribute much to
the ionising flux, but for consistency with the fits by \citet{mehdipour2011} we
include it in our model. However we ignore the weaker narrow and broad X-ray
emission lines that are modelled by \citet{detmers2011} in the soft X-ray band.
Those lines also have a negligible contribution to the ionising flux. More
details about the model components are given in Sect.~\ref{sect:timesed}, where
we discuss the time-dependent parameters of the components.

\begin{table}[!tbp]
\caption{The $^{10}\log$ of the ionising (1--1000 Ryd) flux for the ten
individual spectra 1--10 divided by the corresponding flux for the time-averaged
spectrum, for each of the spectral components as modelled by
\citet{mehdipour2011}.} 
\centering                
\begin{tabular}{@{\,}cccccc@{\,}} 
\hline\hline
Obs & Fe\tablefootmark{a} & PL\tablefootmark{b} & SE\tablefootmark{c} &
DBB\tablefootmark{d} & Total \\
\hline
 1 & $-0.06$& $-0.08$& $-0.07$& $-0.21$&    $-0.08$\\
 2 & $-0.05$& $-0.05$& $-0.06$& $-0.12$&    $-0.06$\\
 3 & $+0.02$& $+0.04$& $+0.01$& $-0.06$&    $+0.02$\\
 4 & $-0.03$& $-0.07$& $+0.07$& $+0.18$&    $+0.03$\\
 5 & $+0.02$& $+0.03$& $+0.12$& $+0.28$&    $+0.09$\\
 6 & $+0.01$& $+0.04$& $+0.02$& $+0.04$&    $+0.02$\\
 7 & $+0.01$& $+0.03$& $-0.01$& $-0.03$&    $+0.01$\\
 8 & $+0.03$& $+0.05$& $-0.05$& $-0.12$&    $-0.02$\\
 9 & $+0.02$& $-0.03$& $+0.00$& $-0.09$&    $-0.01$\\
10 & $+0.01$& $+0.01$& $-0.04$& $-0.15$&    $-0.03$\\
\hline 
\label{tab:deltaxi}                    
\end{tabular}
\tablefoot{
\tablefoottext{a}{Iron line}
\tablefoottext{b}{Power law}
\tablefoottext{c}{Soft excess}
\tablefoottext{d}{Disc blackbody}
}\end{table}

For each of the ten spectra, we have made runs with Cloudy \citep{ferland1998}
version C08.00 with \citet{lodders2009} abundances for a grid of $\xi$-values to
calculate the equilibrium ion concentrations. The auxiliary program
\textsl{xabsinput} in the public SPEX distribution then converts this into an
input file for the \textsl{xabs} model of SPEX. The \textsl{xabs} model was used
in the analysis of the time-averaged RGS spectrum by \citet{detmers2011}. For
each spectrum we also calculate the ionising luminosity $L$ in the 1--1000~Ryd
band (Table~\ref{tab:deltaxi}). 

\subsection{Predicted and observed changes in transmission:
pn\label{sect:transpn}}

\begin{figure*}[!tbp]
\resizebox{0.85\hsize}{!}{\includegraphics[angle=0]{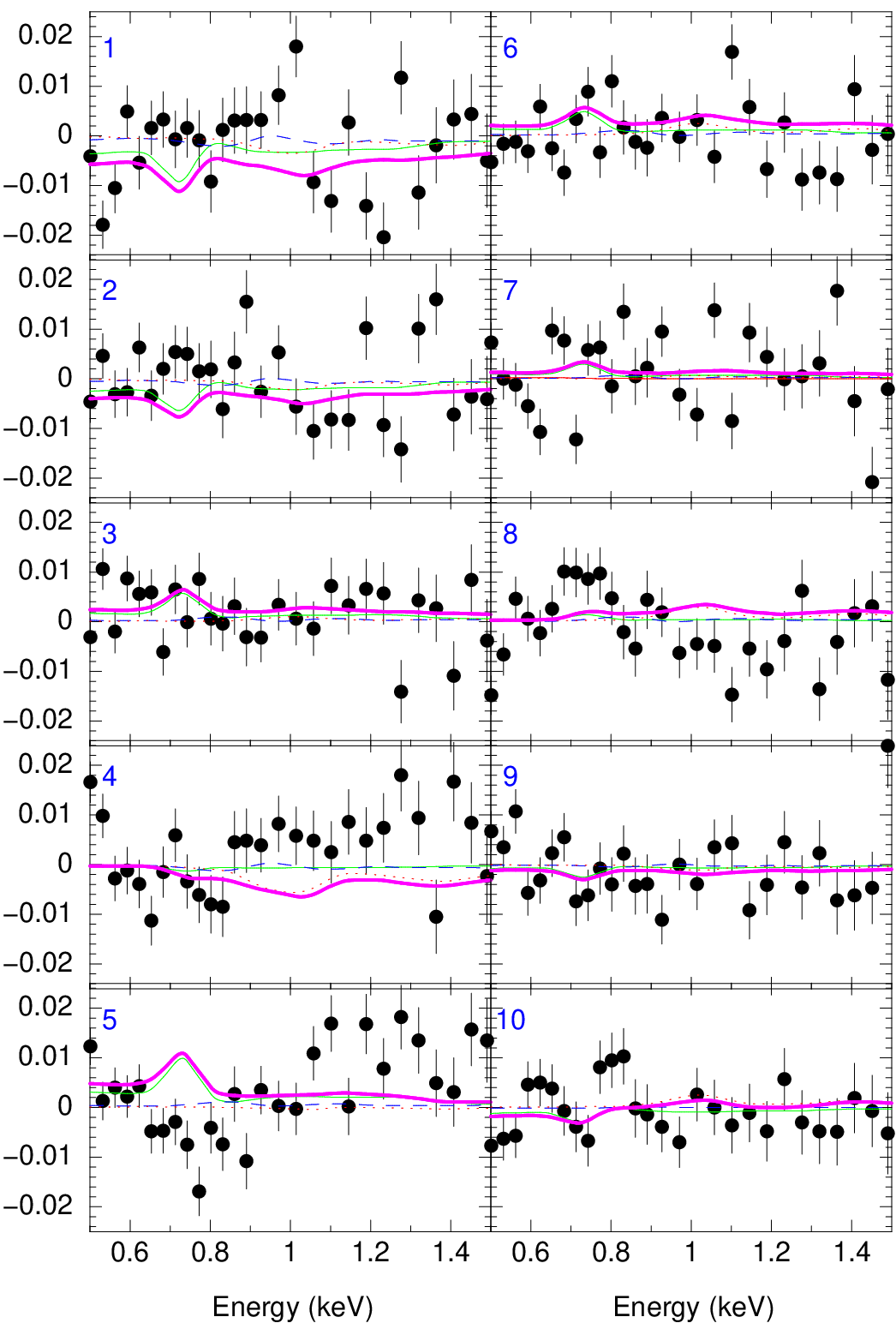}}
\caption{Data points with error bars: observed ratios of the ten spectra 
(labelled 1-10) to the average spectrum in the representation of
Eq.~(\ref{eqn:robs}) (essentially the spectral ratios with the respective
continua divided out and unity-subtracted). Solid green lines: predicted ratio
in the representation of Eq.~(\ref{eqn:rpred}) for  ionisation component C;
dashed blue lines: same for component D; dotted red lines: same for component E;
thick magenta line: components A--E combined. The predictions for components A
and B are close to zero and not shown here separately.
}
\label{fig:delta}
\end{figure*}

As a next step, we calculated for each of the five ionisation components A--E
the transmission using the above appropriate SEDs. We also increased or
decreased  $\xi$ proportionally to the change of the total $L$ from the last
column of Table~\ref{tab:deltaxi}. The ratios of the ten transmissions to the 
time-averaged transmission are then represented using Eq.~\ref{eqn:rpred} and
compared to the observed pn data using Eq.~\ref{eqn:robs}. We show this
comparison in Fig.~\ref{fig:delta}. 

Observations 4 and 5 show some broad-band residuals in the form of a positive
trend between $\sim 0.8$~keV and 1.5~keV, while observation 8 shows the opposite
trend. A similar bending is seen for these spectra in the plot of the ratio of
the model spectra (Fig.~\ref{fig:modelrat}). These three spectra have their
largest curvature just around 1~keV. Thus, the systematic trends in spectra 4, 5
and 8 can be attributed to imperfect continuum modelling. Note that their
amplitude of $\sim$1\% is well below the systematic uncertainty of the pn
effective area calibration. But of course also the continuum models that we use
may be too simple at this level of accuracy.

To corroborate this result, we have inspected the fit residuals of the
broad-band fits to the ten spectra made by \citet{mehdipour2011}. Averaged over
the 1.0--1.5~keV band, the fit residuals of the pn for spectra 4 and 5 are
$\sim$0.8\% higher than for spectra 2, 9 and 10. Similarly, for observation 8
the fit residuals are $\sim$0.4\% smaller than for spectra 2, 9 and 10. Within
0.1\%, these differences agree with what we show in Fig.~\ref{fig:delta}. Thus,
the large-scale residuals are caused by small imperfections in the continuum
modelling of the individual spectra. Recall that the ratios shown in
Fig.~\ref{fig:delta} are essentially the ratios of the observed or predicted
photon spectra, but normalised to the corresponding ratio of the pure continuum
models obtained by \citet{mehdipour2011}. In contrast to these broad-band
trends, the signatures of a varying absorber should be narrow dips or bumps at
specific energies, which differ for each of the absorption components (cf.
Fig.~\ref{fig:transchange}). 

The predicted difference in transmission for \textsl{components A and B} is less
than $\pm$0.02 and $\pm$0.2\%, respectively, and because we are not sensitive
enough to detect such weak signals, we will not consider components A and B
further. We recall that the sensitivity to detecting changes in narrow energy
bands with pn is $\sim$0.5\% or slightly better, cf. Sect.~\ref{sect:epicpn}.

\textsl{Component C} has the strongest predicted signals. Spectrum 1, and to
some lesser extent also spectrum 2, should have a dip with an amplitude of
$\sim$$-$1\% at 0.73~keV, while spectrum 5 should have a positive excess with
similar amplitude. None of this is visible in the data. However, the predicted
dip for spectrum 1 might be visible in spectrum 5, while the predicted bump for
spectrum 5 might be visible in spectrum 8. That would correspond to a delay of
about two weeks. The possibility of a delayed response, corresponding to lower
density and hence more distant gas will be elaborated further in
Sect.~\ref{sect:timedependent}, but we conclude here that the pn data are
clearly inconsistent with an immediate response of component C to continuum
variations.

The predicted changes for \textsl{component D} are too small to be detectable
with pn. This is somewhat surprising because from Fig.~\ref{fig:transchange} one
might conclude that the predicted variations should have about half the
amplitude of the variations in component C. However, in
Fig.~\ref{fig:transchange} it was assumed that the luminosity variations are
achromatic, while in the present model we take the more realistic continuum
shape variations into account. Because the soft X-ray flux varies more strongly
than the hard X-ray flux, the predicted amplitudes for the more highly ionised
components D and E are smaller than that of component C. The strongest predicted
dip, $-0.35$\% around 0.82~keV for spectrum 1, is comparable to the statistical
uncertainty of 0.3\% in that band, hence not detectable.

For \textsl{component E}, our model predicts a dip of $-1.6$\% for spectrum 4 at
1.06~keV, in contrast to a positive peak at the same energy of $\sim +0.5$\% for
spectra 6, 8, and 10. There is not much evidence of the dip in spectrum 4,
although some care should be taken because of the small systematic trends
noticed earlier for this spectrum, related to the strong curvature of the
underlying continuum. 

\subsubsection{Nominal significance of the variations}

\begin{table}[!tbp]
\caption{Comparison of the 10 spectra with models for the change in transmission} 
\centering                
\begin{tabular}{@{\,}c|r@{\,}r@{\,}r@{\,}r|r@{\,}r@{\,}r@{\,}r@{\,}} 
\hline\hline
&\multicolumn{4}{c|}{Original data}&\multicolumn{4}{c}{Corrected data}\\ 
Obs & $\chi^2$\tablefootmark{a} & 
      $\Delta\chi^2$    \tablefootmark{b} & 
      $\Delta\chi^2$    \tablefootmark{b} & 
      $\Delta\chi^2$    \tablefootmark{b} &  
      $\chi^2$\tablefootmark{a} & 
      $\Delta\chi^2$    \tablefootmark{b} & 
      $\Delta\chi^2$    \tablefootmark{b} & 
      $\Delta\chi^2$    \tablefootmark{b} \\ 
    &                           & 
                     (C)                  & 
                     (D)                  & 
                     (E)                  &  
                                & 
                     (C)                  & 
                     (D)                  & 
                     (E)                  \\ 
\hline
 1&    60.9&     2.5&  $-$3.1&     2.5&    56.8&    12.2&  $-$1.2&     3.8\\
 2&    42.3&     8.4&  $-$0.1&  $-$0.6&    39.8&     7.5&     0.0&     0.0\\
 3&    35.9&  $-$3.5&  $-$1.2&  $-$0.5&    32.0&     4.6&     0.3&     0.4\\
 4&    59.1&     1.8&     1.4&    17.8&    29.5&     1.3&     0.0&    10.5\\
 5&    80.4&    19.4&  $-$0.7&     1.0&    33.9&    21.6&     0.7&     0.4\\
 6&    37.3&     1.7&  $-$0.8&  $-$0.3&    34.3&     3.9&  $-$0.2&     0.4\\
 7&    65.7&  $-$0.8&  $-$1.0&  $-$0.4&    63.6&     1.5&  $-$0.5&     0.0\\
 8&    58.9&  $-$1.1&     1.0&     7.2&    31.9&  $-$0.9&     0.6&     4.6\\
 9&    39.8&  $-$0.8&  $-$0.1&  $-$1.0&    31.3&  $-$0.2&     0.1&  $-$0.2\\
10&    27.9&  $-$2.2&  $-$0.1&     2.8&    21.4&  $-$0.9&     0.1&     2.1\\
$\Sigma$  &   508.0&    25.4&  $-$4.8&    28.6&   374.5&    50.4&  $-$0.2&    22.0\\
\hline 
\label{tab:deltchi}                    
\end{tabular}
\tablefoot{
\tablefoottext{a}{$\chi^2$ of the zero model}
\tablefoottext{b}{$\chi^2$ of the prediction for the component indicated in
brackets minus $\chi^2$ of the zero model}
}\end{table}

The significance of our findings is somewhat difficult to assess quantitatively,
due to the small systematic continuum deviations mentioned earlier. We have
attempted to quantify this in the following way (Table~\ref{tab:deltchi}). We
calculated the nominal $\chi^2$ of the data $R_{\rm{obs}}$ shown in
Fig.~\ref{fig:delta} compared to two models. The first model (labelled ``zero
model'') in the table compares the data to the model where the change in
transmission is zero ($R_{\rm{pred}}=0$ in terms of Eq.~\ref{eqn:rpred}). This
corresponds formally to the limit for the gas density $n\rightarrow 0$. We also
calculated $\chi^2$ for the models with instantaneous response for components C,
D, and E (the green, blue and red curves in Fig.~\ref{fig:delta}). This
corresponds formally to $n\rightarrow \infty$. For comparison with the zero
model we subtract $\chi^2$ for that model, thus showing the difference in
$\chi^2$. We do this  for each individual observation. The table clearly shows
that the model with no response at all is better than the models with an
instantaneous response of the absorption components. The numbers agree with what
we described qualitatively before. While the sensitivity to detecting changes in
component D is low, an instantaneous response for component C would make the
total $\chi^2$ 25.4 higher than the zero model. An almost similar conclusion can
be drawn for component E.

\begin{figure}[!tbp]
\resizebox{\hsize}{!}{\includegraphics[angle=-90]{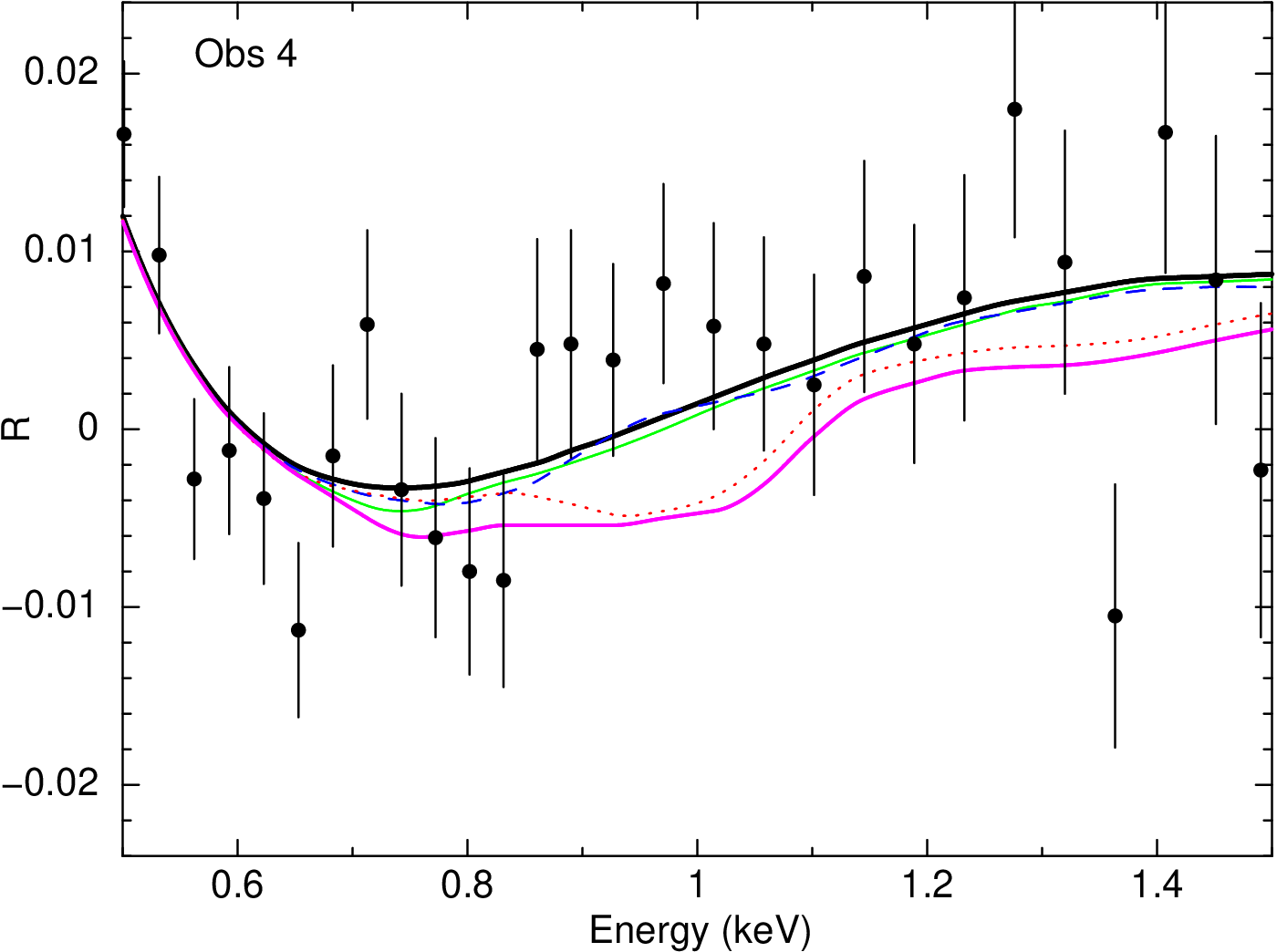}}
\resizebox{\hsize}{!}{\includegraphics[angle=-90]{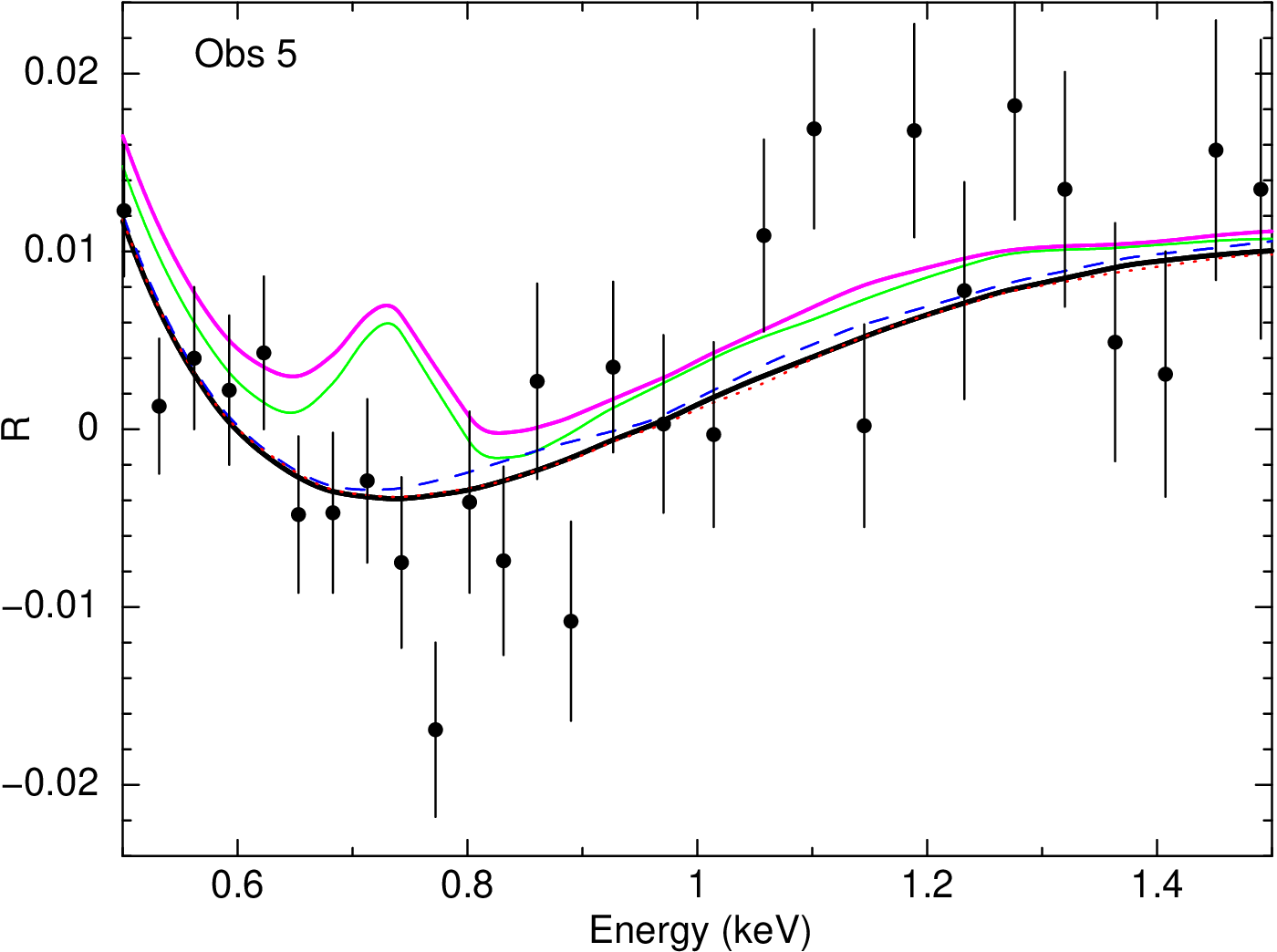}}
\caption{Example of continuum adjustment for spectra 4 (top panel)  and
5 (bottom panel). Data points with error bars: observed ratio of spectra 4 and 5
to the average spectrum in the representation of Eq.~(\ref{eqn:robs}), same as
in Fig.~\ref{fig:delta}.  Solid black line: cubic polynomial fit to $R$ over the
0.4--3.2~keV band as described in the text. Solid green line: predicted $R$ in
the representation of Eq.~(\ref{eqn:rpred}) for  component C, but added here to
the continuum approximation given by the black line; dashed blue line: same for
component D; dotted red line: same for component E; thick magenta line:
components A--E combined.}
\label{fig:ydelta05}
\end{figure}

To test how these results depend on the systematic uncertainties in the
continuum, we re-computed these $\chi^2$ values using corrected data (the
rightmost four columns of Table~\ref{tab:deltchi}). The correction consists of a
cubic polynomial in $\log E$ that we fitted to the data between 0.4--3.2~keV,
thus with a typical resolution of a factor of 2 in energy. An example of such a
fit for spectrum 5 is shown as the black line in Fig.~\ref{fig:ydelta05}. The
correction is not unique, and there is some danger that it washes out a small
part of any true variations of the absorber. Keeping this in mind,  the
correction lowers the overall values of $\chi^2$ compared to the uncorrected
case as expected. It does not alter our conclusion about poorer fits with models
with an instantaneous response compared to models with no response. Thus, we
confirm our more qualitative conclusions given before.

\subsection{Predicted and observed changes in transmission:
RGS\label{sect:transrgs}}

At energies below 1~keV, RGS becomes more sensitive to the detection of weak
lines than pn, thanks to its higher spectral resolution that compensates for the
lower effective area. Therefore we also consider the predicted instantaneous
response to continuum variations for the absorption components as measured by
RGS. We consider only features where the RGS has enough sensitivity to detect
them (Fig.~\ref{fig:rgssens}).

\begin{figure}[!tbp]
\resizebox{\hsize}{!}{\includegraphics[angle=-90]{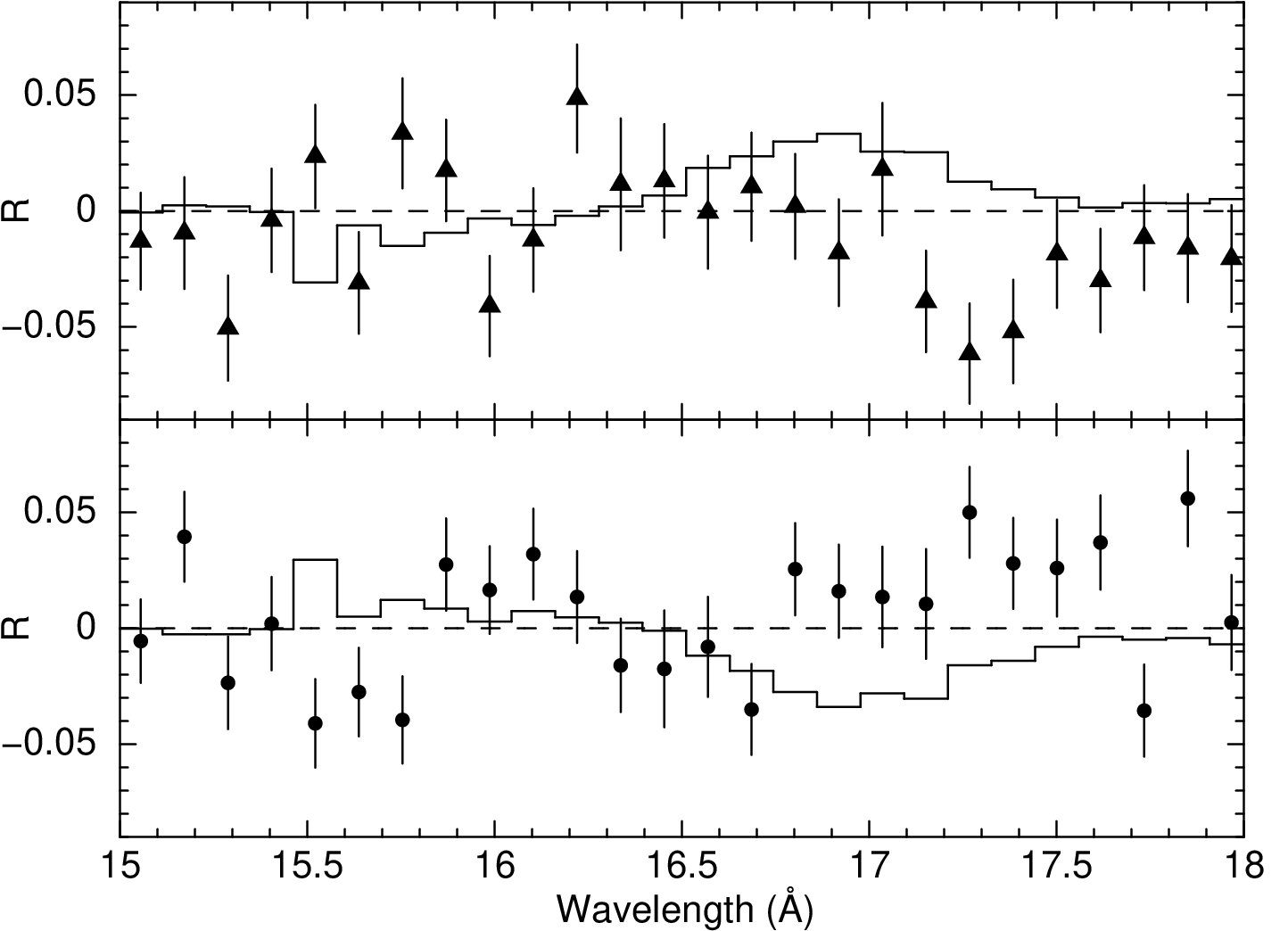}}
\caption{Scaled flux ratios compared to the time-average spectrum
(eqs.~\ref{eqn:robs}--\ref{eqn:rpred}) for RGS. Upper panel:
data and model for spectrum 5; Lower panel: same for the
combined spectra 1 \& 2. The data have been rebinned by another factor of two to
0.116~\AA\ for clarity of display. The models are for component C only.
For comparison with other figures, the band that is plotted corresponds to
0.69--0.83~keV.}
\label{fig:rgs_cvar}
\end{figure}

As for the pn, variations in components A and B are below the detectability
limit so are not considered further. For component C, the 15--18~\AA\ band is
most promising for detecting changes (Fig.~\ref{fig:rgs_cvar}). Our model
predicts the strongest contrast between spectra 1 and 2 on one side, which have
almost the same transmission, and spectrum 5 that corresponds to the peak of the
soft X-ray emission.

Two prominent features contribute to the difference shown in
Fig.~\ref{fig:rgs_cvar}. First, the model predicts a strong change in a broad
band between 16.5--17.5~\AA. This is the same blend as the predicted blend in
the pn data around 0.73~keV (Fig.~\ref{fig:delta}). The main contributions are
from to M-shell transitions of \ion{Fe}{ix}--\ion{Fe}{xiii}, cf.
Table~\ref{tab:line}, and the \ion{O}{vii} edge. Averaged over the
16.5--17.5~\AA\ band, the model predicts an increase of $R$ by $+$0.043 from
observations 1 \& 2 to observation 5. The data show a decrease instead of
$-0.028\pm 0.011$. Thus, at the 6.5$\sigma$ confidence limit we can exclude an
instantaneous response for component C.

The second feature is the \ion{Fe}{xvii} resonance absorption line
(rest-wavelength 15.01~\AA, observed at 15.5~\AA\ due to the cosmological
redshift). Averaged over two adjacent bins, the model for component C predicts a
decrease of $-0.060$ from observations 1 \& 2 to observation 5, while the data
shows a slight increase of $+0.05\pm 0.03$. A part of this increase can be
explained by contamination by component D, which also produces \ion{Fe}{xvii}.
Component D alone produces an increase of $+0.024$. Because the contributions
from components C and D to the total \ion{Fe}{xvii} column density are 59 and
41\%, respectively \citep[cf. Table~4 in][]{detmers2011}, the combined predicted
net decrease in the \ion{Fe}{xvii} line is $-0.026$. This is off by $2.5\sigma$
from the observed value. Thus, the \ion{Fe}{xvii} line is also not in good
agreement with an instantaneous response. 

For component D the largest changes are expected between spectrum 1 and 5, for
the \ion{O}{viii} Ly$\alpha$ line at 19.6~\AA. Averaged over two bins, the
observed increase of $0.02\pm 0.05$ agrees with the prediction of $+0.04$, but
of course also with the hypothesis of no changes. 

Finally, for component E there is a group of features in the 11.3--12.8~\AA\
band where the model predicts a significant change, from a value of $-0.015$ for
spectrum 4 to an average of $+0.007$ for observations 6, 8, and 10. This
increase by $+0.022$ is significantly higher (at the 2.5$\sigma$ level) than the
observed difference of $-0.003\pm 0.010$. Alternatively, fitting a linear
function to the observed $R$-values for all ten observations to the predicted
values yields a slope of $+0.18\pm 0.40$, $2\sigma$ off from the expected value
of 1 for an immediate response. This strengthens our conclusion from the pn data
that also component E does not show much evidence of an instantaneous response.

In the next section we investigate whether the observations agree with a lack of
response or with a delayed response. For this, we need to solve the
time-dependent ionisation balance equations.

\section{Time-dependent photoionisation\label{sect:timedependent}}

\subsection{Introduction}

The time evolution of the ion concentrations in the outflow depends on the gas
density. For high density, the gas responds fast to changes in the ionising flux
and restores ionisation equilibrium at the new flux level. For low density,
collisions between free electrons and ions are rare, and because of that it
takes a long time to restore equilibrium. When the photoionising flux varies
faster than this response time, the plasma may be continuously out of
equilibrium and it is at best in equilibrium with the long-term average flux, on
time scales longer than the response time of the plasma. See
\citet[][Sect.~6]{krolik1995} for a more extensive discussion of these effects.

For strong enhancements of the flux, the ionisation may proceed faster than
recombination \citep[see Fig.~5 of][for a good example of this]{nicastro1999}, 
but when the flux does not vary by orders of magnitude, as in the case of
Mrk~509, the ionisation and recombination time scales are of the same order of
magnitude. 

In this section we present how we model the time-dependent ionisation following
the method of \citet{nicastro1999}. We first determine the history of the
ionising spectral energy distribution (SED) for our observations. Starting with
the first Swift observation, for which we assume ionisation equilibrium, we
calculate for each outflow component the time evolution of the ion
concentrations, for a grid of values for the gas density. We then calculate the
corresponding X-ray transmission for the different components, and compare those
to the observed spectra. The density for each component (or a lower- or upper
limit) is then determined from the best match of a model spectrum to the
observed spectrum.

\subsection{Time-dependent SEDs\label{sect:timesed}}

As in Sect.~\ref{sect:ionisingcontinuum}, we use the detailed spectral fits of
the ten XMM-Newton observations from \citet{mehdipour2011}. They model the
continuum with four components:

\begin{enumerate}

\item disc blackbody (DBB, SPEX model \textsl{dbb}), representing the direct
emission from the accretion disc. Free parameters are the emitting area
$A_{\rm{DBB}}$ and the temperature $T_{\rm{DBB}}$;

\item A soft X-ray excess (SE, SPEX model \textsl{comt}), representing
Comptonised emission from the disc in a warm plasma. Free parameters are the
seed temperature of the photons (taken to be equal to $T_{\rm{DBB}}$), the
plasma temperature $T_{\rm e}$, optical depth $\tau$, and normalisation
$N_{\rm{comt}}$;

\item power law (PL, SPEX model \textsl{pow}), with free parameters the photon
index $\Gamma$ and the normalisation $N_{\rm{PL}}$;

\item Fe-K line (Fe, SPEX model \textsl{gaus}), approximated by a Gaussian and
characterised by its energy $E_{\rm{Fe}}$,  normalisation $N_{\rm{Fe}}$, and
Gaussian width $\sigma$.

\end{enumerate}

We apply a high-energy cut-off to the spectrum, based on the INTEGRAL data
modelling, as shown in \citet[Fig.~3]{kaastra2011a}. The cut-off is represented
by a multiplicative factor to the power-law component, and the shape of the
cut-off is kept constant in time, because more details are lacking. 

The best-fit parameters for the DBB, SE, and PL components are given in
\citet[Table~6]{mehdipour2011}. The parameters for the iron line are not given
in that paper, but the effects of the iron line on the present analysis are
weak, because its ionising flux is only $\sim 0.5$ and 0.2\% of the ionising
flux of the power-law component and the total ionising flux, respectively. For
simplicity, we assume that its flux is constant at the average value for the ten
XMM-Newton observations, 0.76~ph\,m$^{-2}$\,s$^{-1}$, corresponding to
$2.07\times 10^{35}$~W in the source rest frame. Our continuum model is
therefore described by seven free parameters that vary as a function of time.

Because the ionisation state of the gas also depends on the history of the
ionising flux before the start of the XMM-Newton observations, we also need to
characterise that part of the light curve. For that purpose we have obtained
Swift observations as described by \citet{kaastra2011a} and elaborated by
\citet{mehdipour2011}. While for the ten XMM-Newton observations we had high
signal-to-noise pn spectra and OM observations in all filters to constrain the
SED, for the Swift observations, which had short exposure times, we have in most
cases only UV fluxes for the UVM2 filter and broadband X-ray fluxes at 0.3 and
4~keV with a typical uncertainty of 0.05--0.10~dex \citep[see Figs. 2 and 8
of][]{mehdipour2011}. 

These three fluxes (UV, 0.3 and 4 keV) per Swift observation are in principle
insufficient for deriving the seven continuum parameters for the SED. However,
there are tight correlations between these parameters, and therefore we can make
a good proxy for the SED based on the UVM2 and the 4 keV fluxes alone, as we
show below. We use the higher signal-to-noise and more detailed XMM-Newton
observations to derive these correlations. For the Swift observations without
the UVM2 filter, we estimate the UV flux from the other filters as described by
\citet{mehdipour2011}. 

\begin{figure}[!tbp]
\resizebox{\hsize}{!}{\includegraphics[angle=-90]{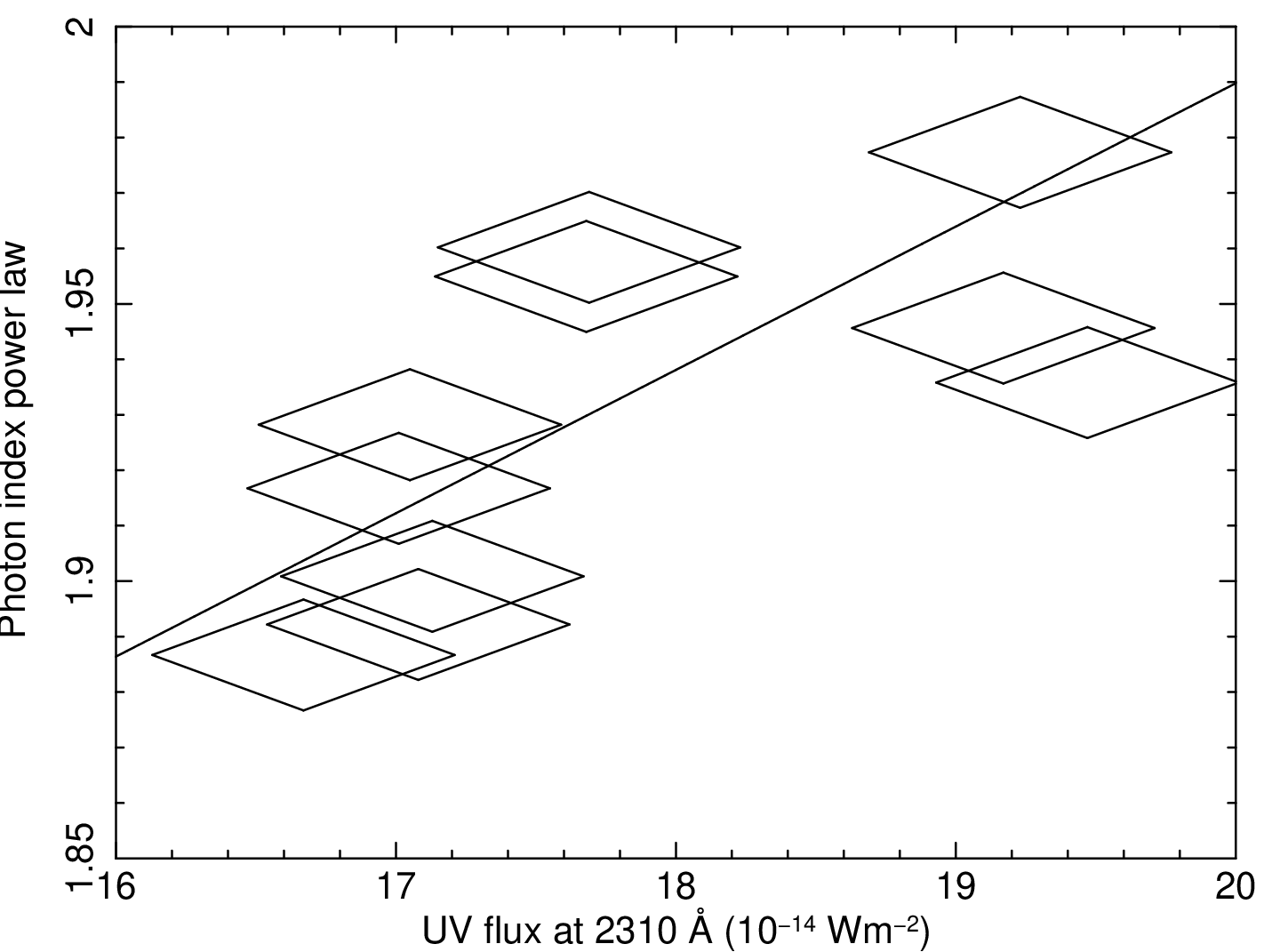}}
\caption{Correlation between the photon index of the PL component with the UV
flux for the ten XMM-Newton observations. The corners of the diamonds correspond
to the nominal statistical uncertainty on both quantities.}
\label{fig:cor1}
\end{figure}

For the {\sl power-law component}, the photon index $\Gamma$  correlates
remarkably well with the UV flux (Fig.~\ref{fig:cor1}). The PL is then
completely determined from the UV flux and the 4~keV X-ray flux, which has no
contribution from the other softer continuum components.

\begin{figure}[!tbp]
\resizebox{\hsize}{!}{\includegraphics[angle=-90]{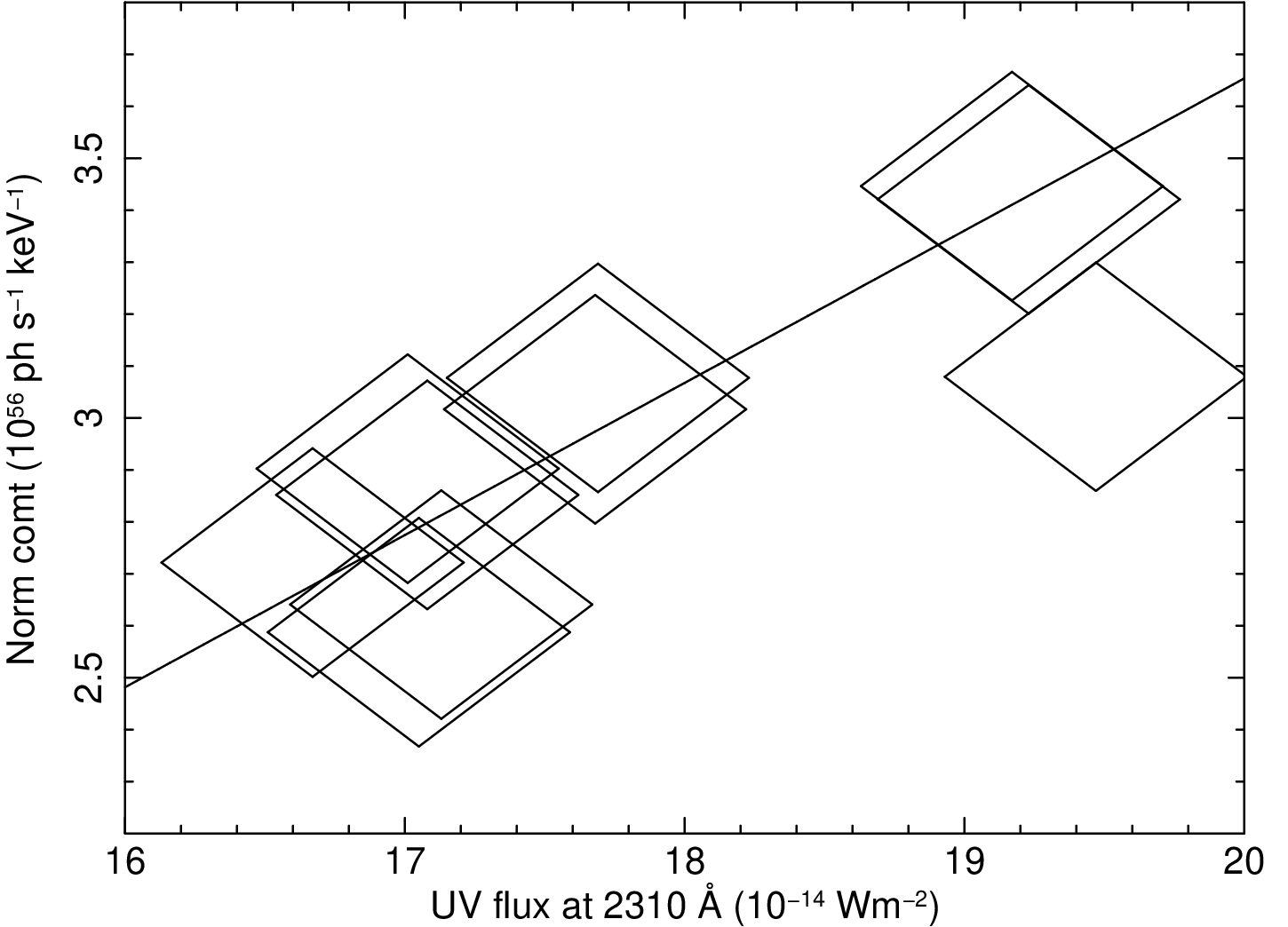}}
\resizebox{\hsize}{!}{\includegraphics[angle=-90]{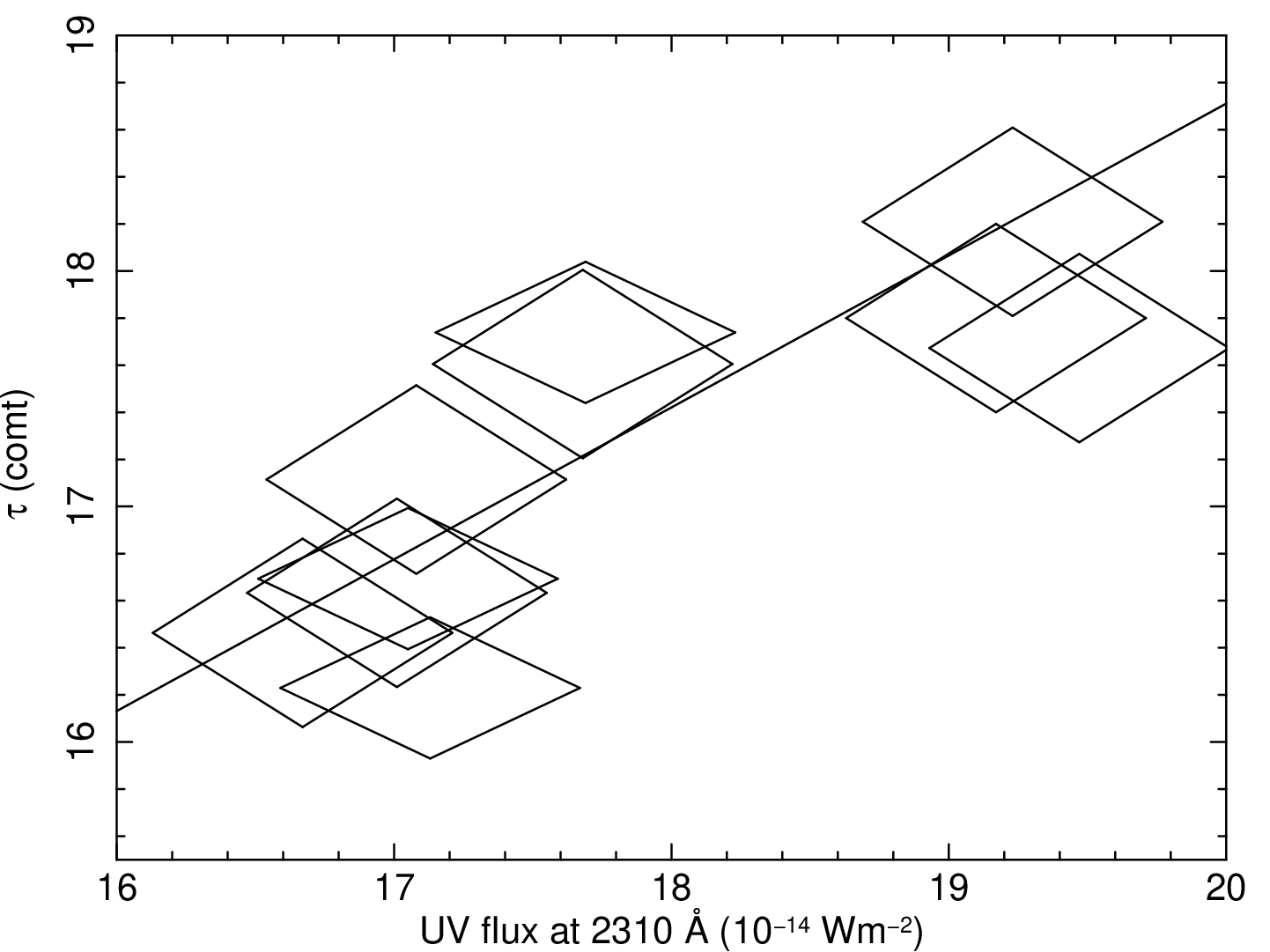}}
\caption{Correlation between the normalisation $N_{\rm{comt}}$
and optical depth $\tau$ of the soft excess (SE) component with the UV
flux for the ten XMM-Newton observations. The corners of the diamonds correspond
to the nominal statistical uncertainty on both quantities.}
\label{fig:cor23}
\end{figure}

For the {\sl soft excess component}, the plasma temperature as given by
\citet{mehdipour2011} shows little variation, and thus we keep it constant at
0.20~keV. The normalisation $N_{\rm{comt}}$ and optical depth $\tau$ of this
component can be approximated as a linear function of the UV flux
(Fig.~\ref{fig:cor23}), and as stated before, the temperature of the seed
photons is tied to the disc blackbody.

\begin{figure}[!tbp]
\resizebox{\hsize}{!}{\includegraphics[angle=-90]{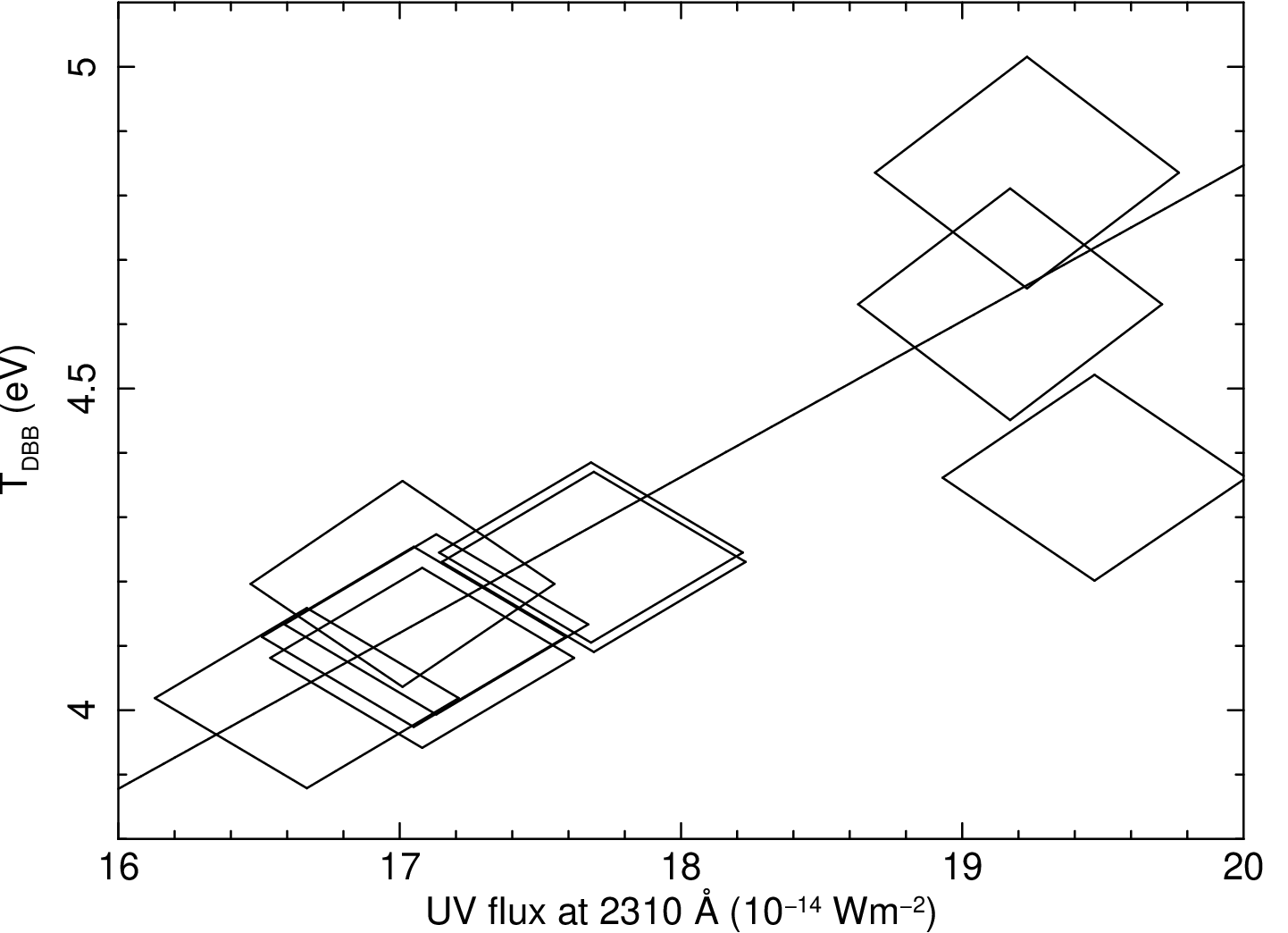}}
\caption{Correlation between the temperature of the DBB component and the UV
flux for the ten XMM-Newton observations.The corners of the diamonds correspond
to the nominal statistical uncertainty on both quantities. }
\label{fig:cor4}
\end{figure}

Finally, for the {\sl disc blackbody component}, there is a correlation between
its temperature and the UV flux (Fig.~\ref{fig:cor4}). We kept its normalisation
constant to $3.34\times 10^{25}$~m$^{2}$, the mean value for the ten
observations, and within the error bars, consistent with the reported values
\citep[Table~6]{mehdipour2011}.

\begin{figure}[!tbp]
\resizebox{\hsize}{!}{\includegraphics[angle=-90]{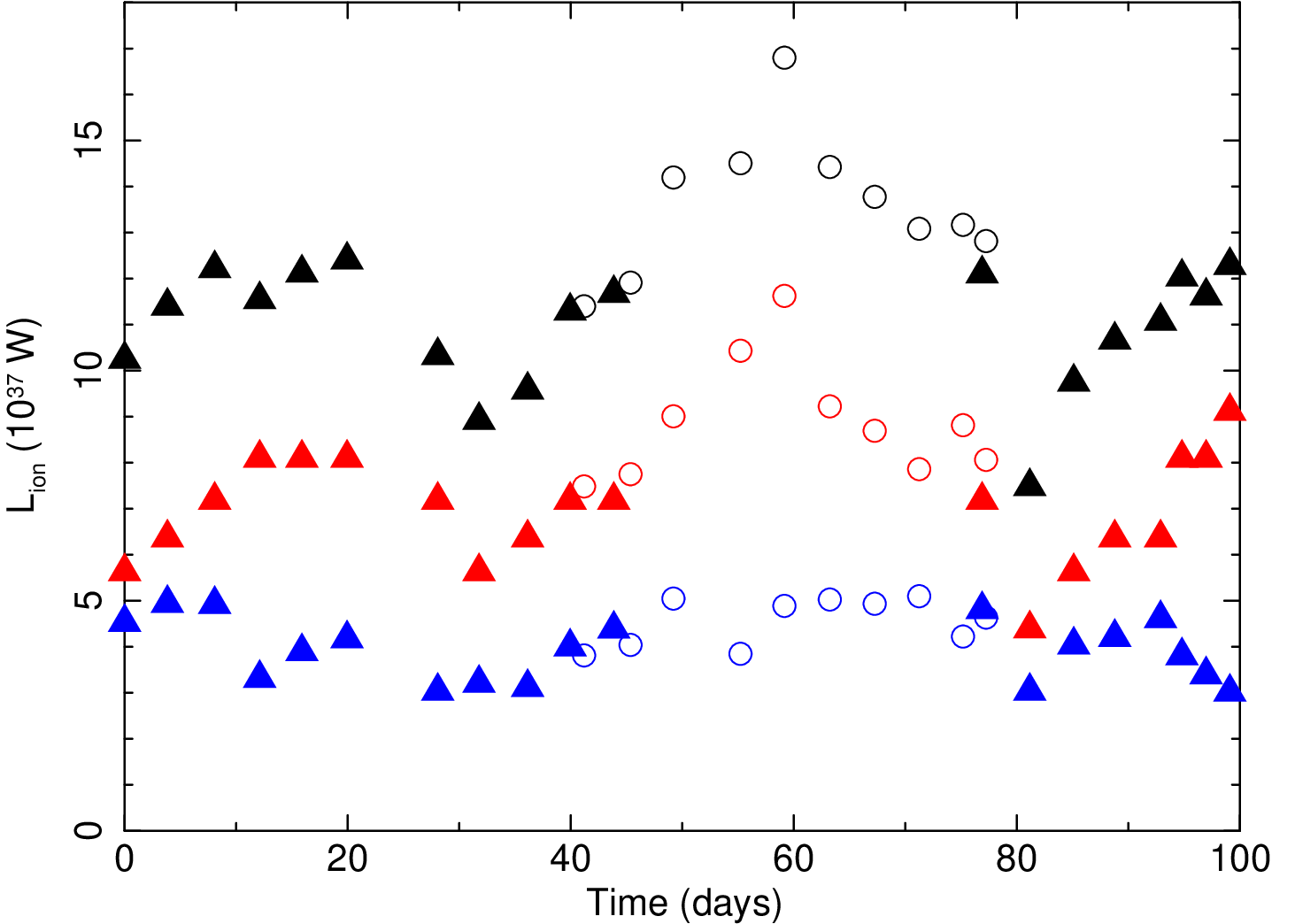}}
\caption{Ionising luminosity (1--1000 Ryd) for the power-law component (blue),
Comptonised component (red), and total spectrum (black). Filled triangles: based
on Swift data; open circles: XMM-Newton observations.}
\label{fig:lumplot}
\end{figure}

Using these correlations and the UVM2 and 4~keV flux, we constructed the
time-dependent SEDs for our full 100 day campaign. Obviously, for the XMM-Newton
observations we took the SEDs directly from \citet{mehdipour2011} and we
reported them in Sect.~\ref{sect:ionisingcontinuum}. A comparison of the
ionising fluxes determined directly from the XMM-Newton data with the ionising
fluxes determined through the above correlations shows a scatter of $\sim
10$\%. 

We show the ionising fluxes of the main components (PL and SE) in
Fig.~\ref{fig:lumplot}. It is seen that the luminosity before and after the
XMM-Newton observations is lower than during these XMM-Newton observations. Most
of the variability is caused by the soft excess component, although the
power-law component also varies. The ionising luminosities of the disc-blackbody
and iron line are much lower than the luminosities of the PL and SE components.

\subsection{Solution of the time-dependent rates}

The time dependence of the relative density $n_i$ of an ion $i$ of a given
species is given by

\begin{equation}
\frac{{\mathrm d}n_i}{{\mathrm d}t} = -n_i n_{\rm e} \alpha_{i-1}
+ n_{i+1} n_{\rm e} \alpha_{i} -n_i I_i + n_{i-1} I_{i-1}.
\label{eqn:ionrate}
\end{equation}
Here $n_{\rm e}$ is the electron density, and $\alpha_i$ is the recombination
rate from stage $i+1$ to stage $i$ and is in general a function of the
temperature $T$. The ionisation rate from stage $i$ to $i+1$ is indicated by
$I_i$. Obviously, in  Eq.~\ref{eqn:ionrate} ionisations from bare nuclei and
recombinations onto neutral atoms should be excluded as they are non-existent.

Equation~(\ref{eqn:ionrate}) forms a set of coupled ordinary differential
equations. We solve these numerically by using a Runge-Kutta method with
adaptive stepsize control, subroutine \textsl{odeint} from \citet{press1992}.
Following \citet{nicastro1999}, we adopt a constant electron density $n_{\rm e}$
as a function of time. 

The recombination and ionisation rates are interpolated linearly between the
values for the 29 different Swift and XMM-Newton observations. We also tried
quadratic or spline interpolation, but that sometimes leads to unwanted
overshooting, in particular near the start and the end of the XMM-observations
where a few Swift observations are close in time to those by XMM-Newton.
Moreover, such higher order schemes would hide the intrinsic uncertainties
caused by the lack of data between the 29 observations.

The recombination and ionisation rates for the 29 epochs were obtained as
follows. For each epoch and ionisation component, we calculated the equilibrium
ion concentrations using Cloudy \citep{ferland1998} version C08.00, with
\citet{lodders2009} abundances. These equilibrium concentrations were calculated
using the appropriate SED (see previous subsection). Furthermore, the ionisation
parameters $\xi$ were scaled down relative to the values obtained from the
average XMM-Newton spectrum \citep{detmers2011}, proportionally to the ionising
1--1000 Ryd luminosity $L$. For a prompt response of the gas to changes in flux
or SED shape, this is the appropriate scaling, assuming that the gas density $n$
and distance to the ionising source $r$ remain constant in time ($\xi=L/nr^2$).
Using this same Cloudy run, we used the ``punch ionization rates'' option of
Cloudy to obtain the corresponding ionisation and recombination rates. Since
Cloudy takes multiple ionisation after inner-shell ionisation into account, the
total ionisation rate obtained from the punch option does not return an
equilibrium situation when inserted into our simplified Eq.~(\ref{eqn:ionrate}),
which ignores multiple ionisations. To correct for this we have scaled the
ionisation rates $I_i$ in our code so that equilibrium is forced (${\mathrm
d}n_i/{\mathrm d}t=0$ in Eq.~\ref{eqn:ionrate}) using the concentrations $n_i$
provided by Cloudy. 

We did these calculations for a grid of distances $r$ to the ionising source
and for each of the outflow components separately. We assumed equilibrium for
the first Swift spectrum as our initial condition.

\subsection{Results}

\subsubsection{An illustrative example}

\begin{figure}[!tbp]
\resizebox{\hsize}{!}{\includegraphics[angle=-90]{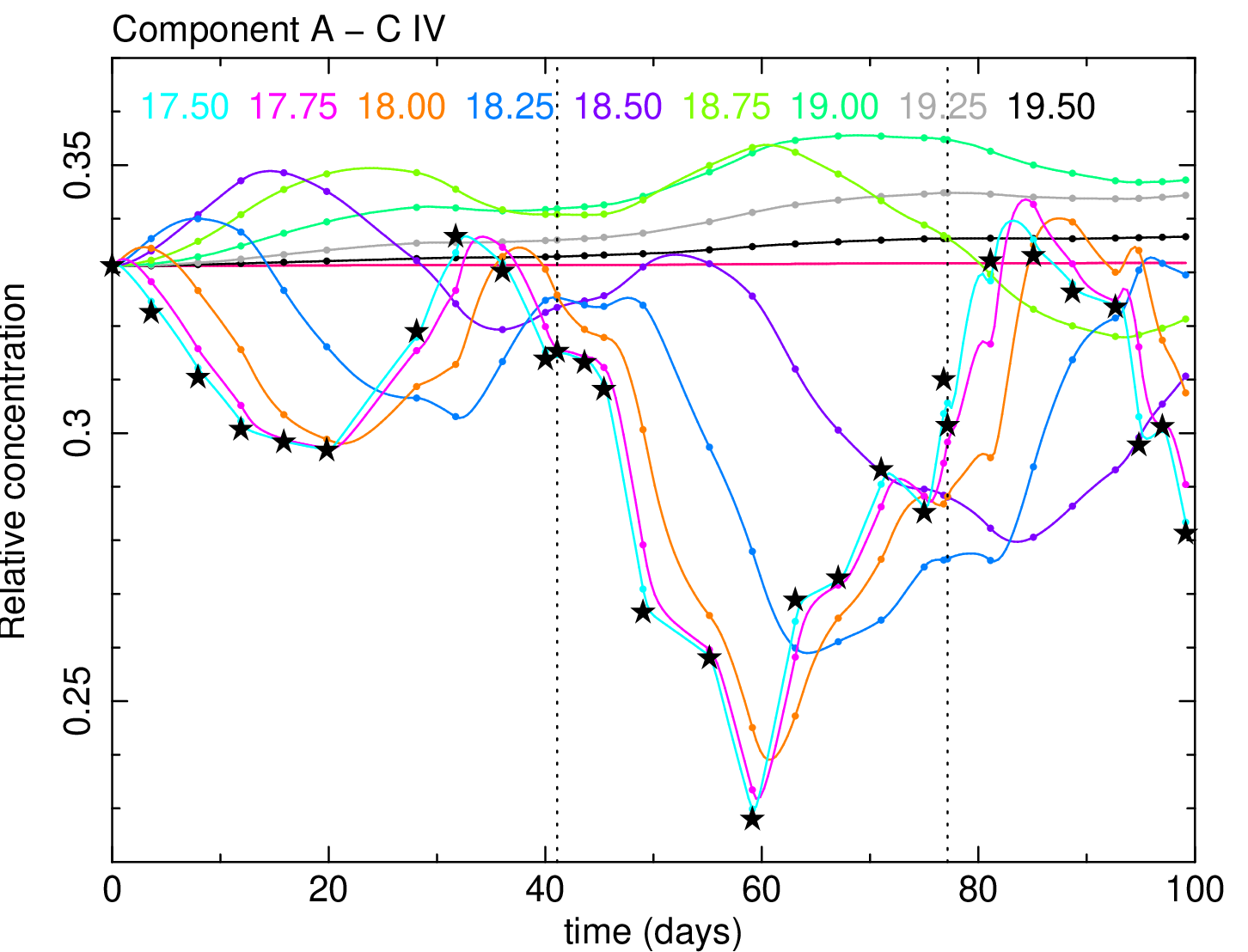}}
\caption{Time evolution of the \ion{C}{iv} concentration of component A,
relative to the total carbon concentration. Stars: a model with instantaneous
response to continuum variations. Coloured lines: predictions for absorbing
material at different distances as indicated by the number near the top of the
plot; these numbers represent the $\log$ of the distance in m. The
XMM-Newton observations were taken between the two vertical dotted lines.}
\label{fig:l_a0604}
\end{figure}

We start here with \ion{C}{iv}, an ion that mainly produces UV absorption but
displays a number of interesting features in the time-dependent ion
concentration. Fig.~\ref{fig:l_a0604} shows the predicted evolution of the ion
concentration as a function of time for the dominant \ion{C}{iv} contributor,
component A. 

For an instantaneous response, the relative \ion{C}{iv} concentration varies
between 0.23 and 0.34 in accordance with the continuum variability. For
distances up to $10^{17.75}$~m, there is not much difference with the
instantaneous response. For distances of $10^{18}$ and $10^{18.25}$~m, the
response becomes delayed by about three and ten days, respectively. These delays
are related to the lower densities and therefore longer recombination time
scales corresponding to the longer distances (for constant $L/\xi=nr^2$).

For greater distances, for instance $10^{18.50}$~m, the delay becomes a month
but also the amplitude of the variations starts to decrease rapidly. The plasma
is no longer able to respond to the fastest fluctuations of the ionising
continuum. For even greater distances ($>10^{18.75}$~m), the delay becomes
longer than the duration of six week of the Swift monitoring before the start of
the XMM-Newton observations, and the amplitude becomes significantly smaller.
Here our model predictions become quite uncertain, as we lack sufficient
information on the flux history of the years before the start of our campaign.
There are only some sparse individual measurements available. 

\subsubsection{Time evolution for the most prominent ions}

\begin{figure*}[!tbp]
\begin{center}
\resizebox{0.42\hsize}{!}{\includegraphics[angle=-90]{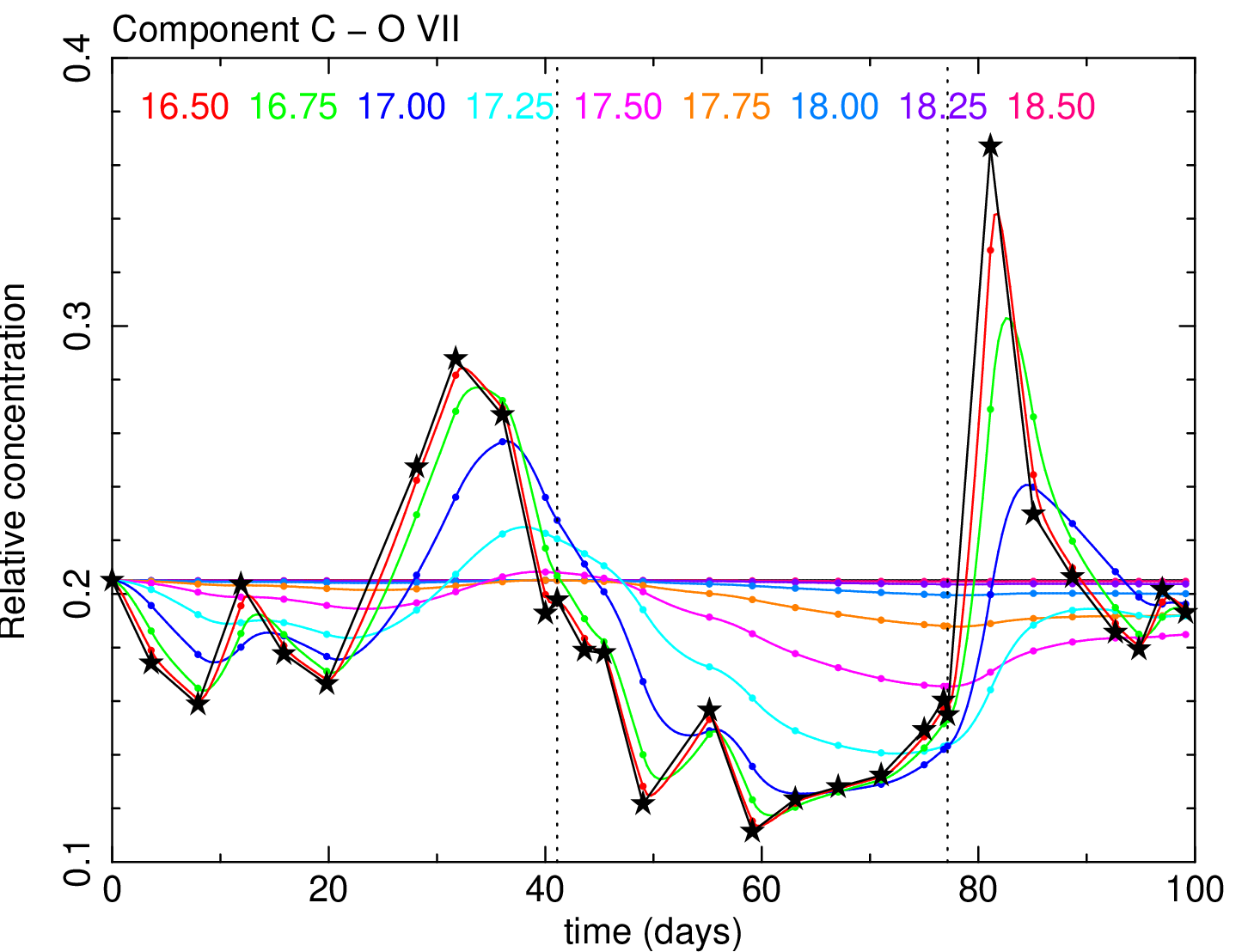}}
\,\,
\resizebox{0.42\hsize}{!}{\includegraphics[angle=-90]{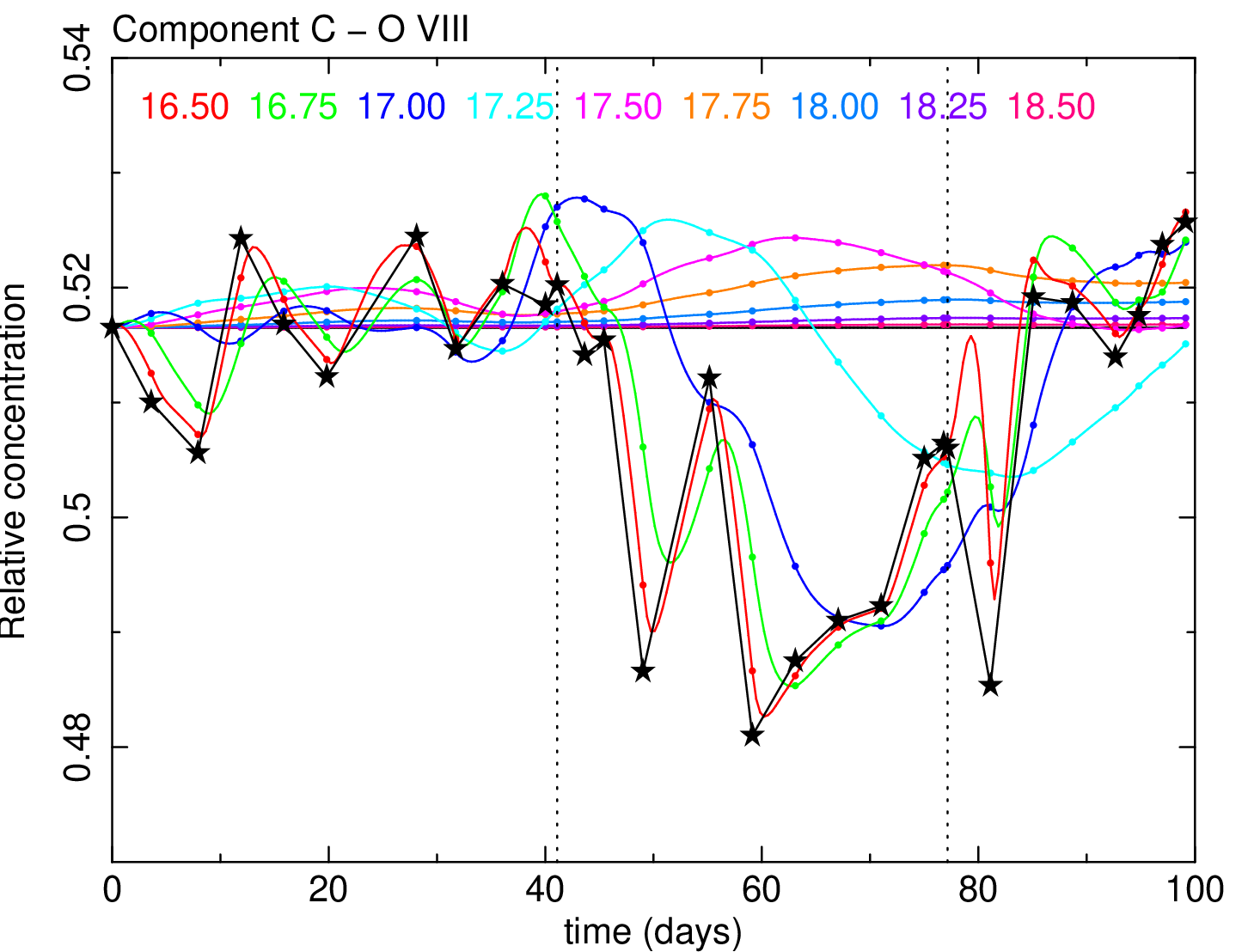}}
\,\,
\resizebox{0.42\hsize}{!}{\includegraphics[angle=-90]{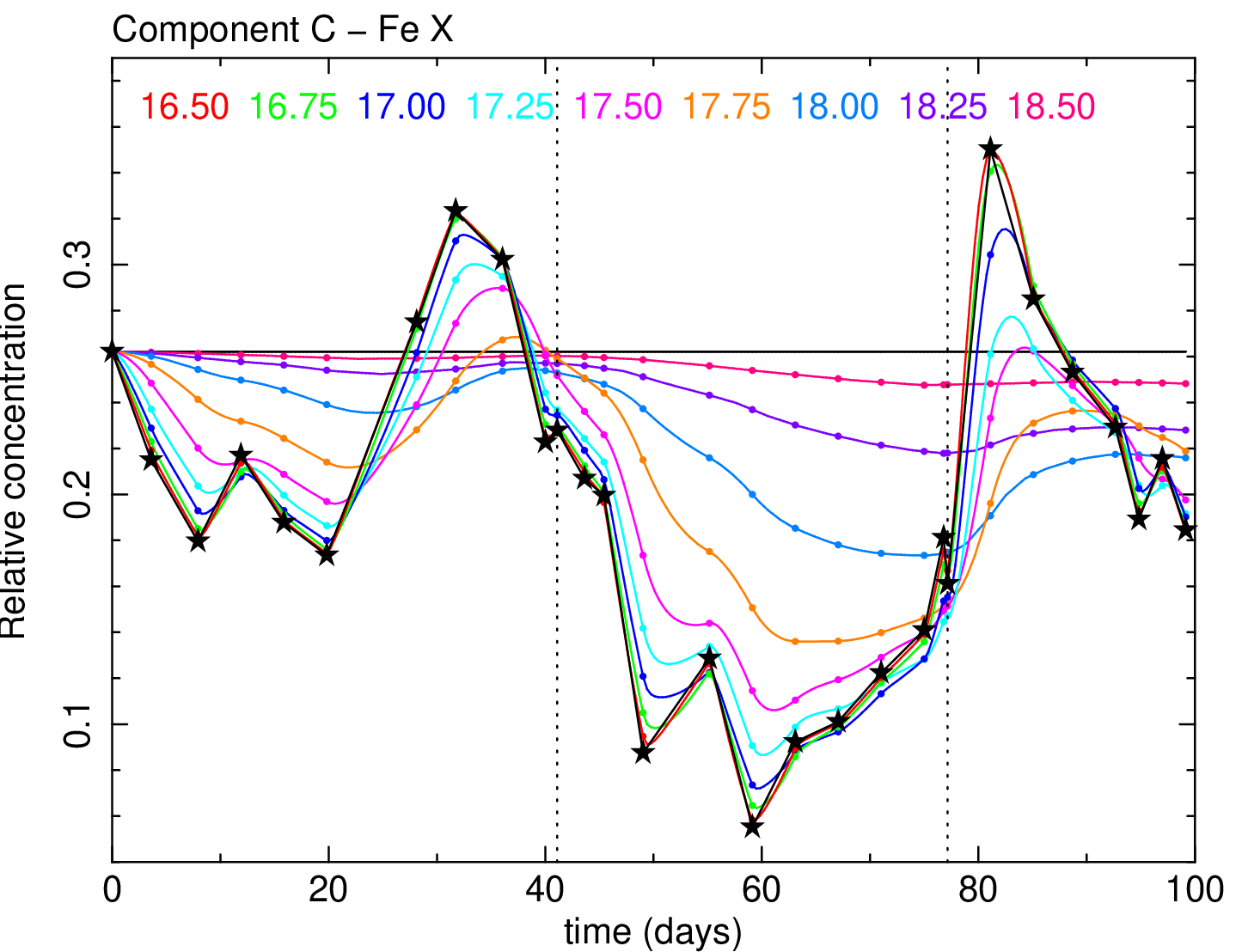}}
\,\,
\resizebox{0.42\hsize}{!}{\includegraphics[angle=-90]{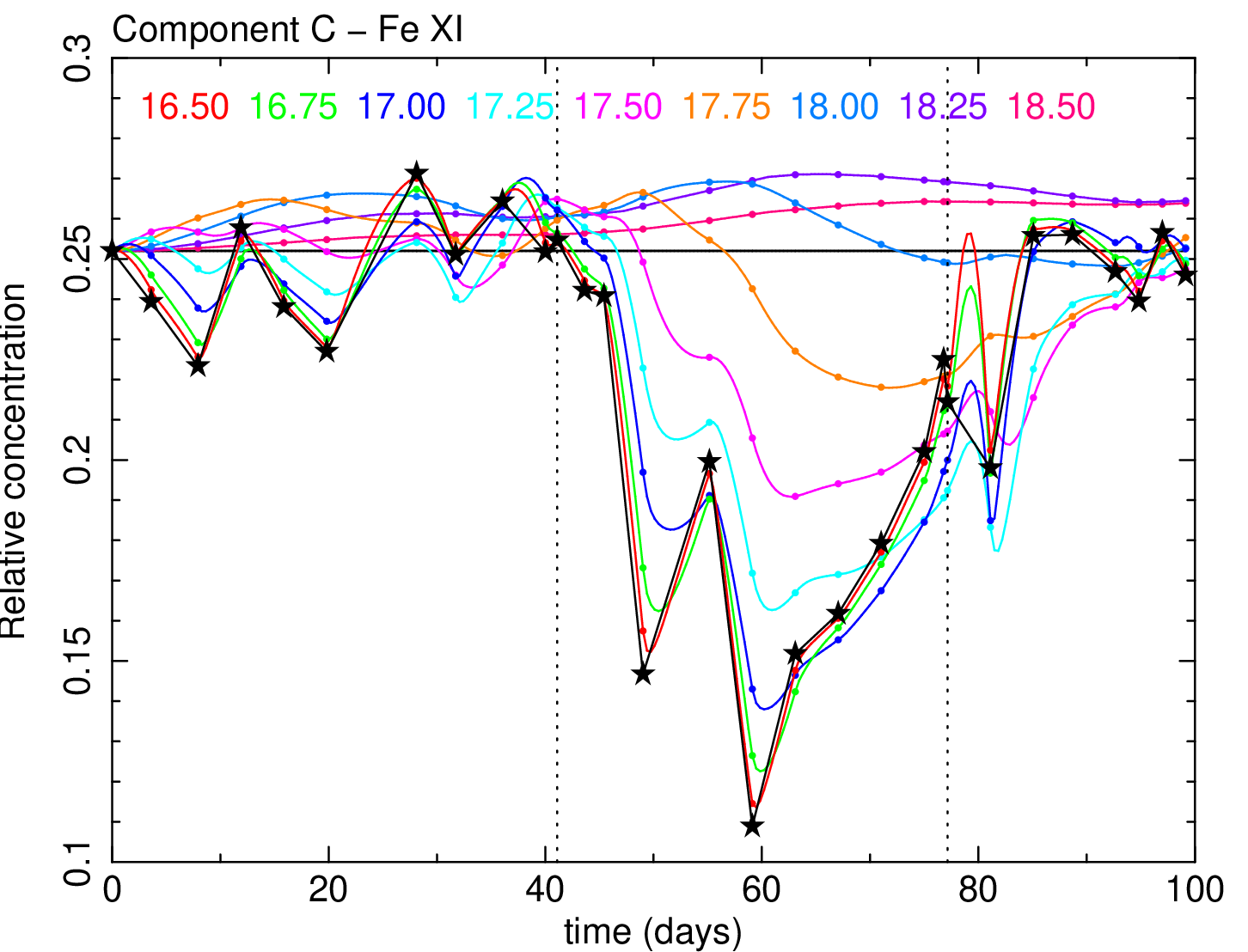}}
\,\,
\resizebox{0.42\hsize}{!}{\includegraphics[angle=-90]{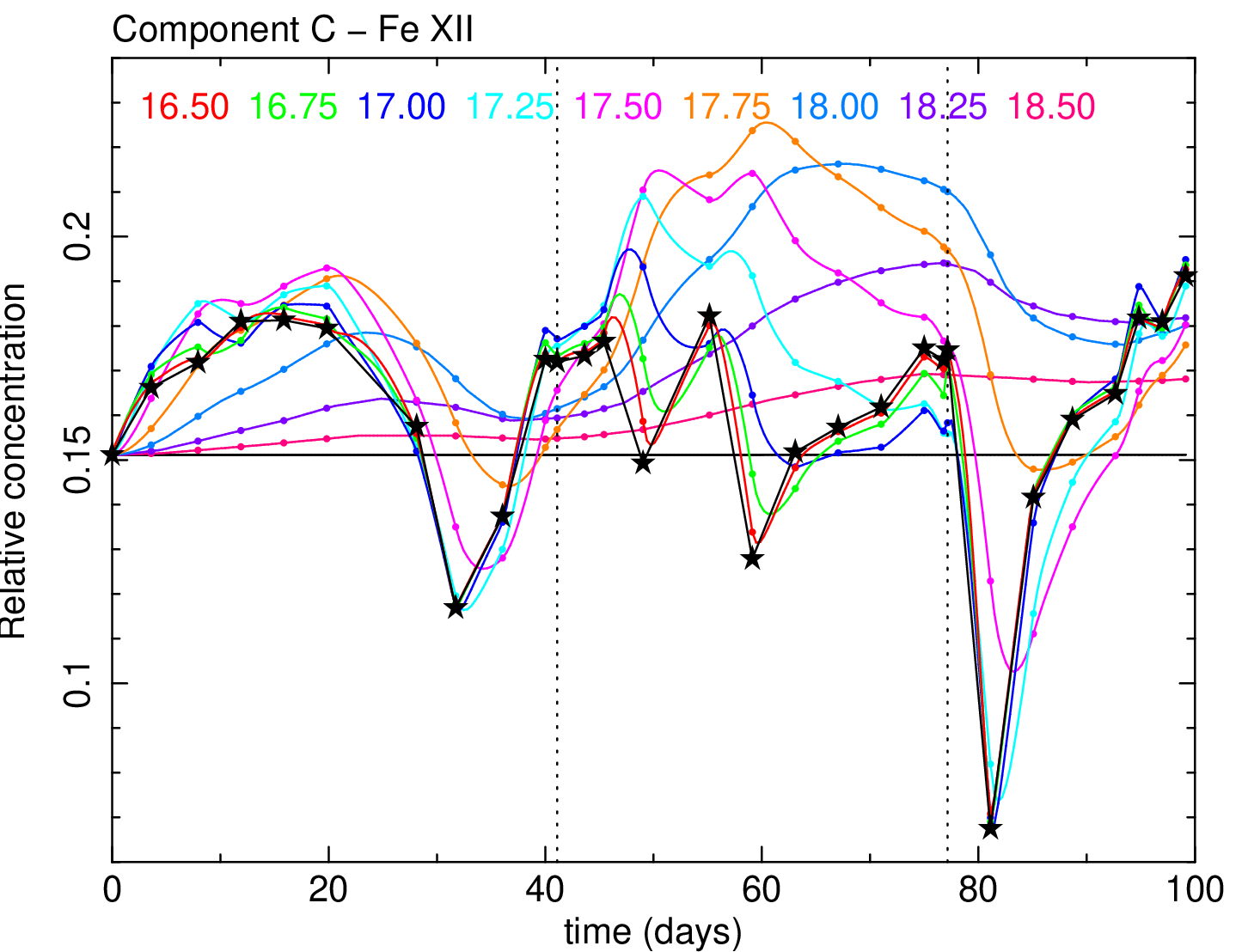}}
\,\,
\resizebox{0.42\hsize}{!}{\includegraphics[angle=-90]{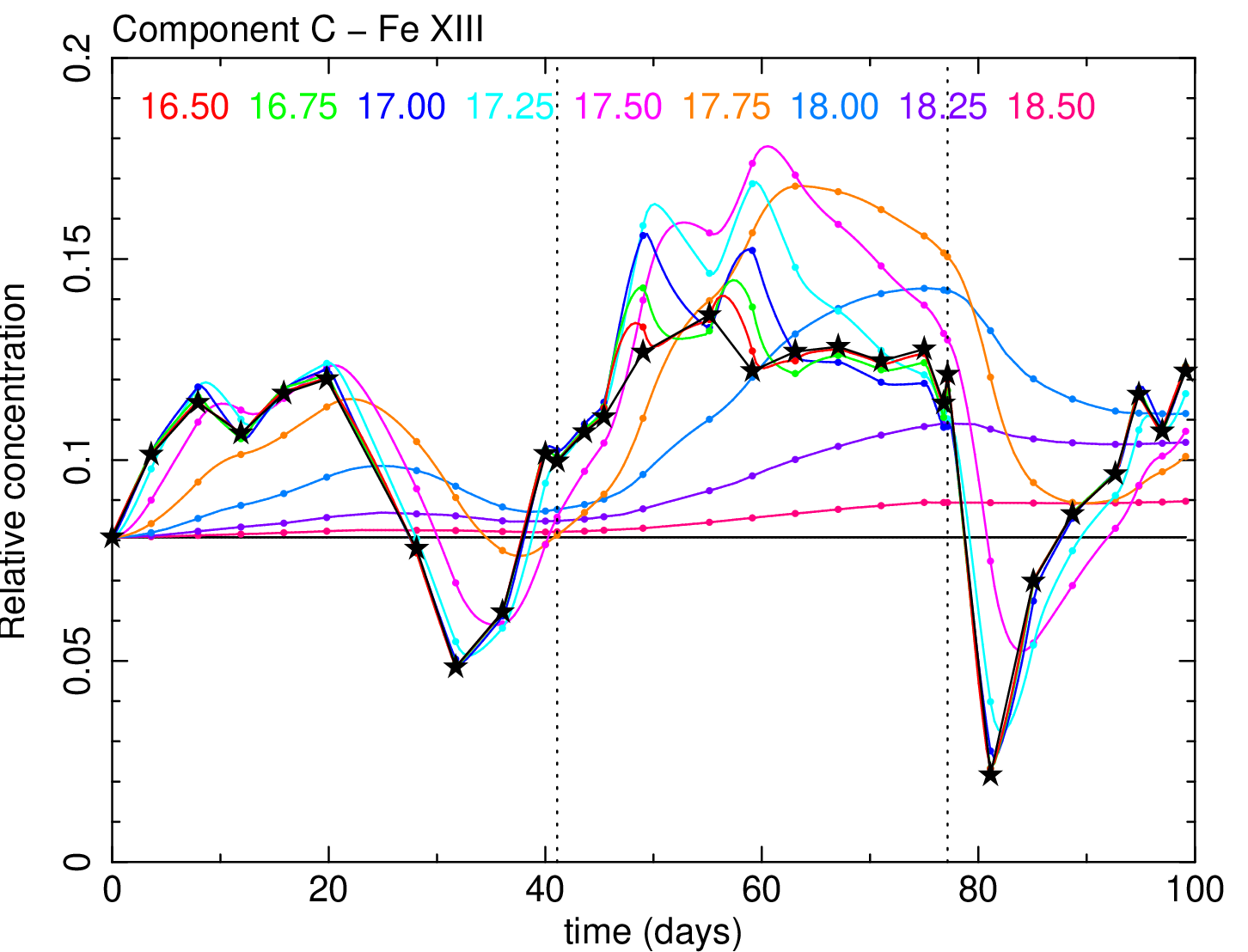}}
\,\,
\resizebox{0.42\hsize}{!}{\includegraphics[angle=-90]{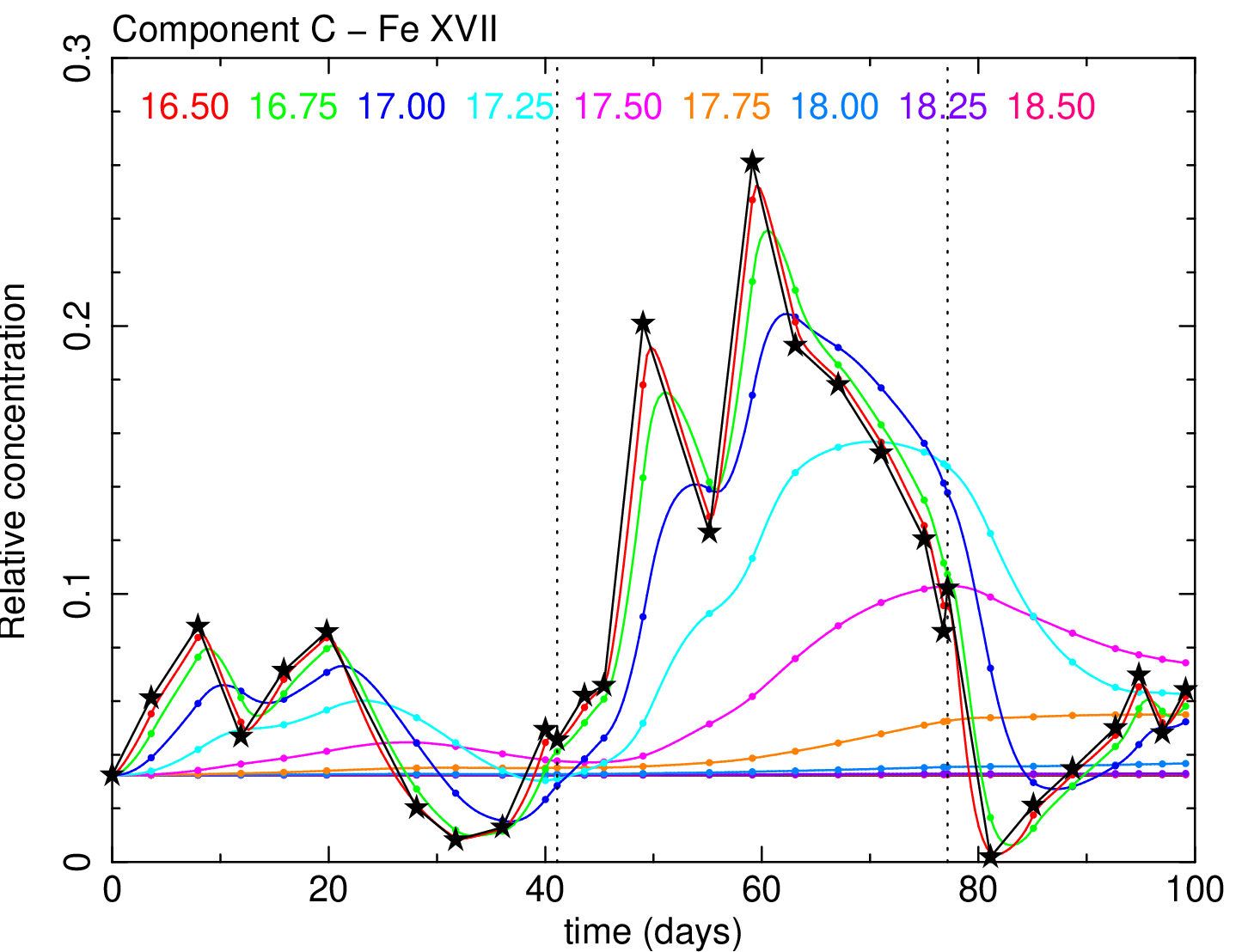}}
\,\,
\resizebox{0.42\hsize}{!}{\includegraphics[angle=-90]{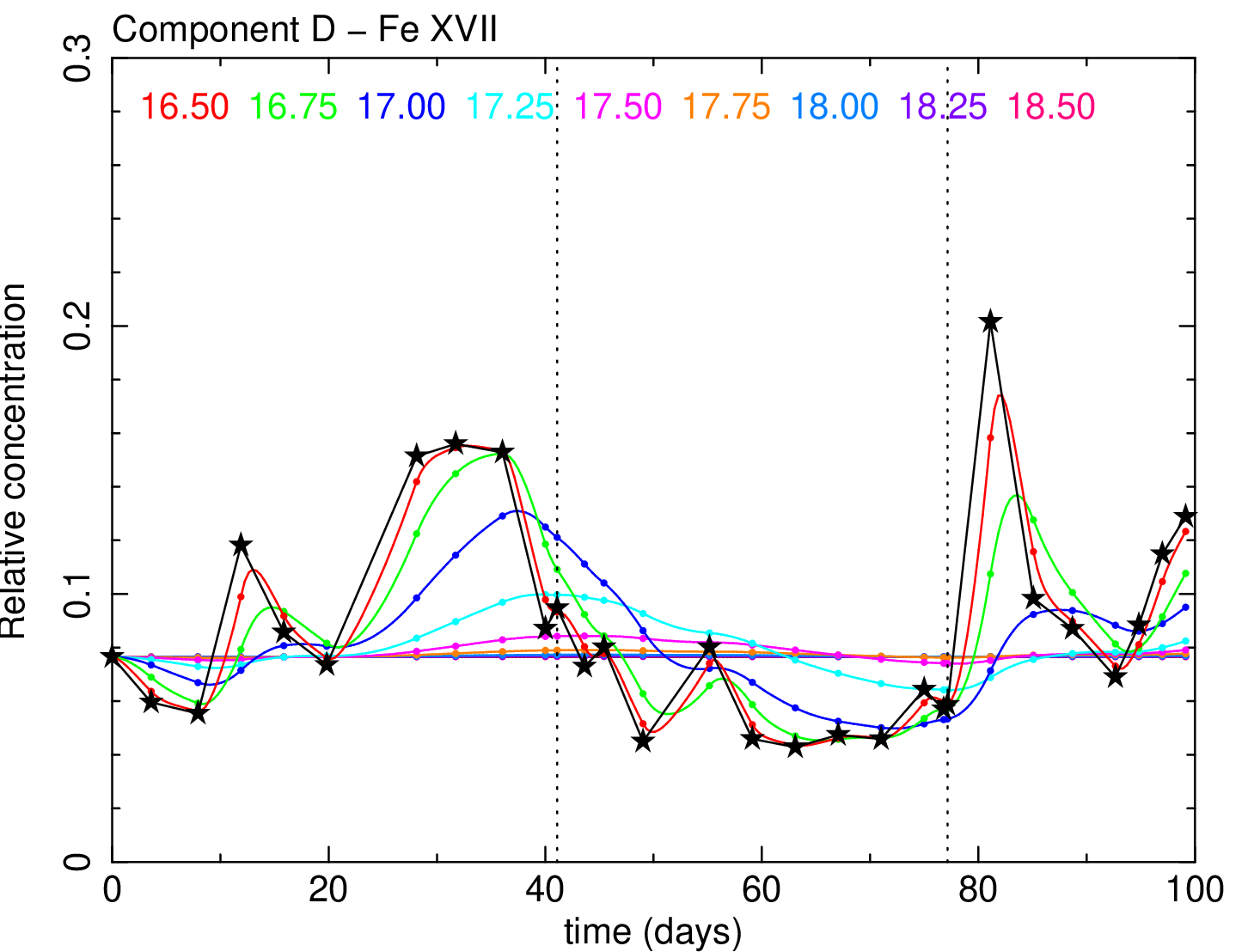}}
\end{center}
\caption{Time evolution of the concentrations of several ions in components
C and D. Further explanation see Fig.~\ref{fig:l_a0604}.}
\label{fig:l_ions}
\end{figure*}

In Fig.~\ref{fig:l_ions} we show the time evolution of the most important ions
for our variability study (see Table~\ref{tab:line}). Starting with oxygen, we
see that the strong soft X-ray flux enhancement near the centre of the
XMM-Newton observations has a strong ionising effect: both the \ion{O}{vii} and
\ion{O}{viii} concentrations decrease for gas at distances $<10^{17.25}$~m; for
longer distances, the history on longer time scales becomes more important.

Comparing \ion{Fe}{x} with \ion{O}{vii}, it is seen that iron responds faster to
continuum changes than oxygen: a similar delay is found for iron at three times
greater distances (compare the curves for \ion{O}{vii} and \ion{Fe}{x} at
$10^{17.00}$ and $10^{17.50}$~m, respectively). For the iron M-shell ions
\ion{Fe}{x} -- \ion{Fe}{xiii} the lower ionisations stages show a decrease in
column density while the higher ionisation stages show an increase near the
centre of the XMM-Newton observations.

Finally, the two panels for \ion{Fe}{xvii} for components C and D show that the
response not only strongly depends on the distance to the central source, but
also on the ionisation parameter: component D has a six times higher ionisation
parameter than component C, and it responds in a less pronounced fashion.

\subsubsection{Changes in the transmission}

With the calculated time-dependent column densities, we next predict the changes
in the transmission for each component as a function of distance. We compare the
calculated transmission for individual observations to the calculated
time-averaged transmission for all ten observations, similar to the spectral
comparisons shown in Fig.~\ref{fig:delta}. This average is different for each
adopted distance, cf. the different curves in Fig.~\ref{fig:l_ions}.

\begin{table}[!tbp]
\caption{Lower limits to the distance of the absorption components
for different energy bands and instruments, at the 90\% and 99\% confidence
levels.} 
\centering                
\begin{tabular}{@{\,}cc@{\,}c@{\,}ccc@{\,}} 
\hline\hline
     &         &       &                 &  90\% lim.  & 99\% lim. \\
Com- & Instru- & Range & $\chi^2$/d.o.f. & $\log r$ & $\log r$ \\
ponent & ment  &       &                 & (m)      & (m) \\
\hline
C    & pn & 0.5--1.5~keV   & 508/279   & $>$18.26 & $>$17.99 \\
C    & pn & 0.62--0.78~keV & 109/59    & $>$17.84 & $>$17.62 \\
C    & RGS & 7--36~\AA\    & 7056/4972 & $>$18.29 & $>$18.11 \\
C    & RGS & 16.5--17.5~\AA\ & 217/169 & $>$18.32 & $>$18.11 \\
C    & RGS & 15.4--15.6~\AA\ & 45/19   & $>$17.58 & $>$17.37 \\
D    & RGS & 19.5--19.7~\AA\ & 33/19   & $>$17.16 & $>$17.02 \\
E    & pn  & 0.5--1.5~keV    & 508/279 & $>$17.15 & $>$17.01 \\
E    & RGS & 11.25--12.8~\AA\ & 342/259& $>$16.50 & $>$16.42 \\
\hline 
\label{tab:chitest}                    
\end{tabular}
\end{table}

In Sect.~\ref{sect:denslim} we found no evidence of an instantaneous response
for any of the components where our sensitivity was sufficient to detect it.
Here we follow a more general approach. For each ionisation component and for a
grid of distances, we determine the formal agreement of the predicted ratio
$R_{\rm{pred}}$ with the observed $R_{\rm{obs}}$ by calculating $\chi^2$-values
in given energy or wavelength bands. We summarise our findings in
Table~\ref{tab:chitest}. In all cases, the best agreement was obtained for the
assumption of no response (the low-density limit), so our data provide lower
limits to the distances of the absorption components. The table also lists the
$\chi^2$-value and degrees of freedom for the low-density limit (no response).
The offsets in $\chi^2$-values from the degrees of freedom are caused by the
slight mismatch of our continuum model for some of the observations (see
Sect.~\ref{sect:transpn}) but these do not affect our distance estimates because
they depend on the relative increases in $\chi^2$ and not on the absolute
values. 

For component C, the comparison of the pn results for the 0.5--1.5 and
0.62--0.78~keV band shows that the highest sensitivity is reached by using the
broader band. This is in line with Fig.~\ref{fig:delta}, which shows that, apart
from the main peak at 0.72~keV, there is also a weaker contribution from the
broader 1.0--1.5~keV band. The RGS data in the 16.5--17.5~\AA\ band yield a
distance limit that is almost equal to the limit obtained from the pn data. The
narrower band around \ion{Fe}{xvii} (15.4--15.6~\AA) also yields lower limits,
but these are less restrictive than the limits from the broader band.

For component D, the pn data cannot distinguish the different cases well so we
only get a lower limit from the \ion{O}{viii} Ly$\alpha$ line as measured by
RGS. For component E the pn spectrum provides the best limit.

\begin{table}[!tbp]
\caption{Combined lower limits to the distance and density of the absorption
 components at the 90\% confidence level.} 
\centering                
\begin{tabular}{@{\,}cccccc@{\,}} 
\hline\hline
Component & Distance & Density \\
          & (pc)     & (10$^9$~m$^{-3}$)\\
\hline
C    & $>$71  & $<$0.28 \\
D    & $>$4.7 & $<$10.6\\
E    & $>$4.6 & $<$1.7 \\
\hline 
\label{tab:distance}                    
\end{tabular}
\end{table}

Combining our results for component C from pn and RGS, and taking the most
restrictive limits for components D (RGS) and E (pn), Table~\ref{tab:distance}
summarises the 90\% confidence lower limits to the distance and density of
components C--E.

\section{Long-term variability\label{sect:longterm}}

\begin{figure}[!tbp]
\resizebox{\hsize}{!}{\includegraphics[angle=0]{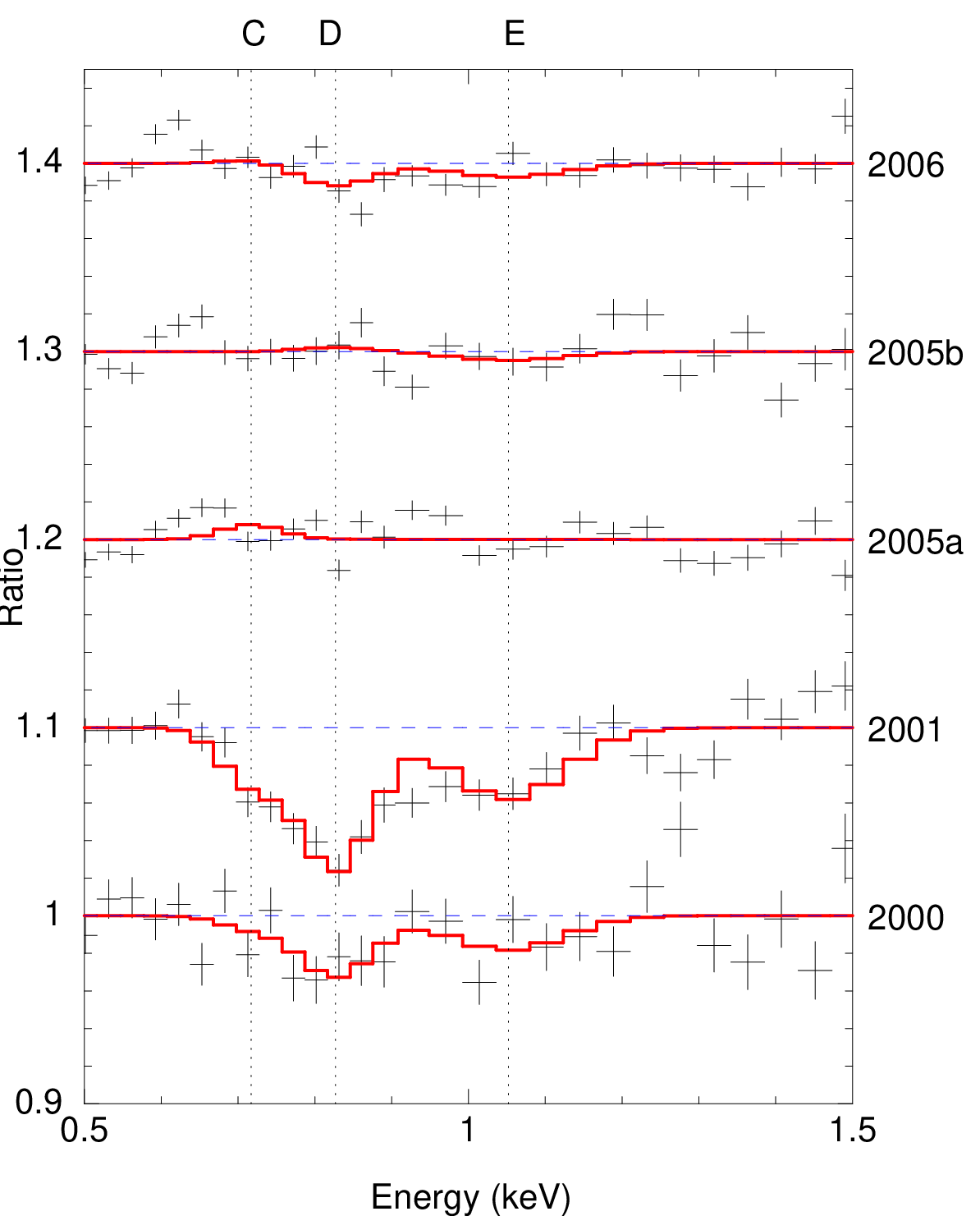}}
\caption{Observed ratios of the fluxed pn spectra of Mrk~509 taken between 2000
and 2006 to the stacked spectrum taken in 2009, in the 0.5--1.5 keV band. As
described in the text, a local power-law approximation to the continuum ratios
has been taken out. Dotted vertical lines indicate the centroids for three
Gaussian components (Table~\ref{tab:gausblend}) that represent the expected changes
of the transmission associated to the ionising flux differences. The solid
curves indicate the best-fit sum of three Gaussians to the residuals. For
clarity of display, the curves for each spectrum have been shifted along the
y-axis by multiples of 0.1.}
\label{fig:longterm}
\end{figure}

\begin{table}[!tbp]
\caption{Parameters for the model of a power law plus three Gaussians fitted to
the pn spectra of Mrk~509 taken between 2000--2006, divided by the 2009 spectrum
as described in the text.} 
\centering                
\begin{tabular}{@{\,}l@{\enskip}c@{\enskip}c@{\enskip}c@{\enskip}c@{\enskip}c@{\enskip}c@{\,}} 
\hline\hline
yyyy-mm-dd & $t_{\rm{exp}}$\tablefootmark{a} & $\Delta\Gamma$\tablefootmark{b} 
& Flux\tablefootmark{c} & $N_{\rm C}$\tablefootmark{d} &
$N_{\rm D}$\tablefootmark{d} & $N_{\rm E}$\tablefootmark{d}\\
\hline
2000-10-25 & 20.4 & 0.296(8) &0.4244(16) & $-$0.3(3) & $-$1.3(4) & $-$0.8(4)\\
2001-04-20 & 30.1 & 0.107(5) &0.6558(17) & $-$1.8(4) & $-$4.9(4) & $-$2.5(4)\\
2005-10-16 & 59.8 & 0.070(3) &0.7483(12) & $+$0.6(3) & $+$0.0(3) & $+$0.0(3)\\
2005-10-18 & 32.4 & 0.006(5) &0.7885(17) & $-$0.0(4) & $+$0.2(4) & $-$0.4(4)\\
2006-04-25 & 47.7 & 0.129(4) &0.7717(14) & $+$0.1(3) & $-$0.9(3) & $-$0.6(3)\\
\hline
\label{tab:longterm}                         
\end{tabular}
\smallskip
\tablefoot{
Digits in parenthesis are the standard deviations in the last digits of the
given value\\
\tablefoottext{a}{Net exposure time (ks) of the pn observation; the
total exposure time for the 2009 observations is 430.6~ks}
\tablefoottext{b}{Power-law index of the ratio; more positive numbers indicate
harder spectra}
\tablefoottext{c}{Ratio of the normalisations at 1~keV}
\tablefoottext{d}{Normalisations of the Gaussians for components C--E as
labelled, multiplied by 100.}
}
\end{table}   

In the previous section we have shown that there are no indications of a change
in the absorption components C--E during our intensive XMM-Newton monitoring
period of 36 days. Here we investigate the longer-term variability. To this aim
we have investigated the five archival XMM-Newton observations taken between
2000--2006. Similarly to the analysis of the data obtained during our campaign,
we divided the fluxed pn spectra for all these observations by the fluxed pn
spectrum of the combined 2009 observations. As a first step, we simply fit the
ratios of the fluxed spectra to a power law plus three Gaussians in the 0.5--1.5
band. For the Gaussians we keep the centroids and widths fixed to the values
given in Table~\ref{tab:gausblend}, representing the expected differences due to
changes in the ionising luminosity. We plot these ratios, with the local
power-law model divided out, in Fig.~\ref{fig:longterm}. We list the best-fit
amplitudes for the Gaussians in Table~\ref{tab:longterm}.

It is clear from Fig.~\ref{fig:longterm} and Table~\ref{tab:longterm} that there
are no significant differences in transmission between the  2005 and 2009
spectra. There may be a hint for a small difference for component D in the 2006
spectrum, but it is not very strong.

The strongest differences are visible in the 2000 and 2001 spectra. For that
reason, we have done a more sophisticated analysis of the continuum spectrum for
those two observations. We also extracted the XMM-Newton OM fluxes for those
observations and did spectral fits to both spectra using exactly the same model
as \citet{mehdipour2011}; see also Sect.~\ref{sect:timesed} for a description of
the spectral model they used. 

\begin{figure}[!tbp]
\resizebox{\hsize}{!}{\includegraphics[angle=-90]{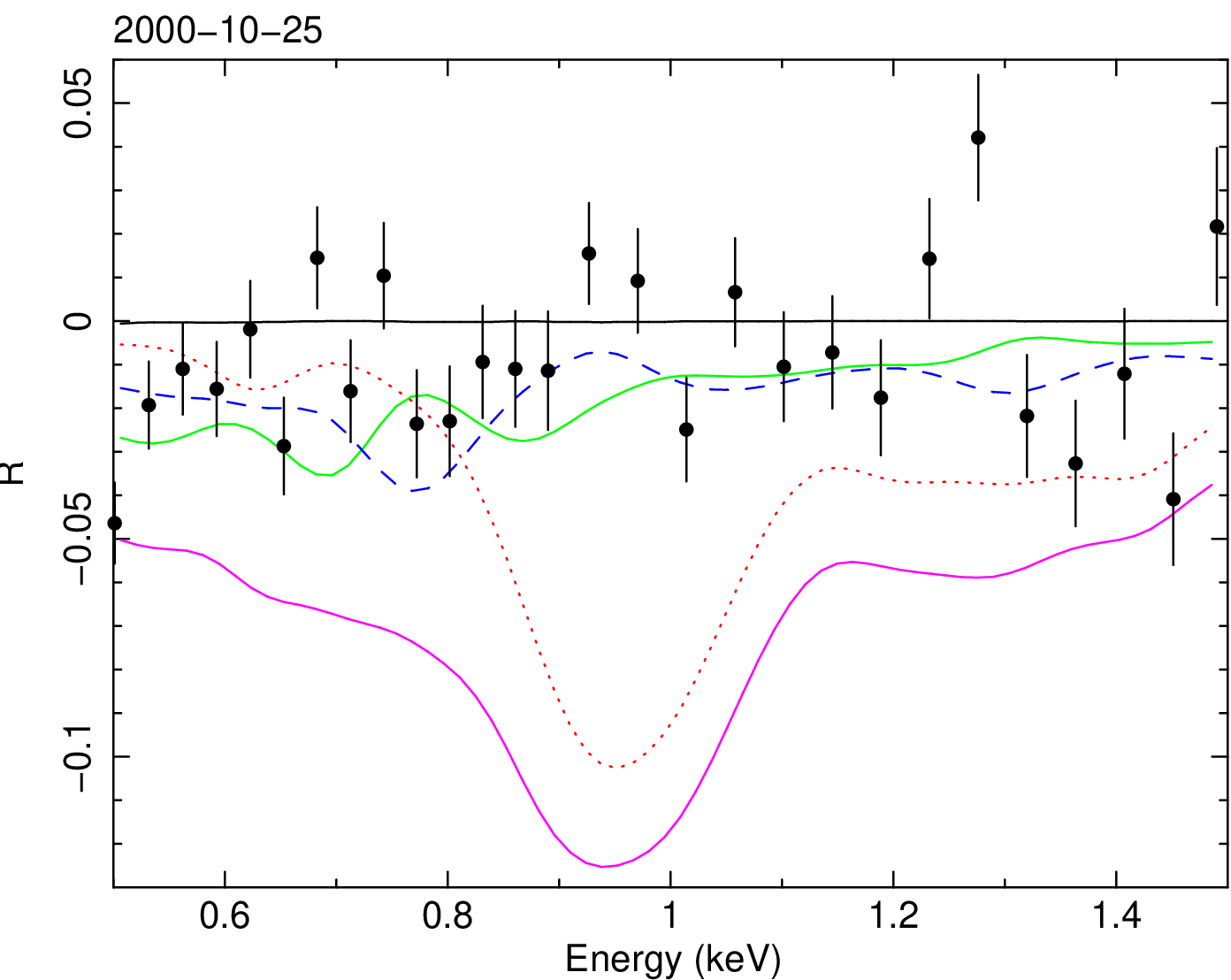}}
\resizebox{\hsize}{!}{\includegraphics[angle=-90]{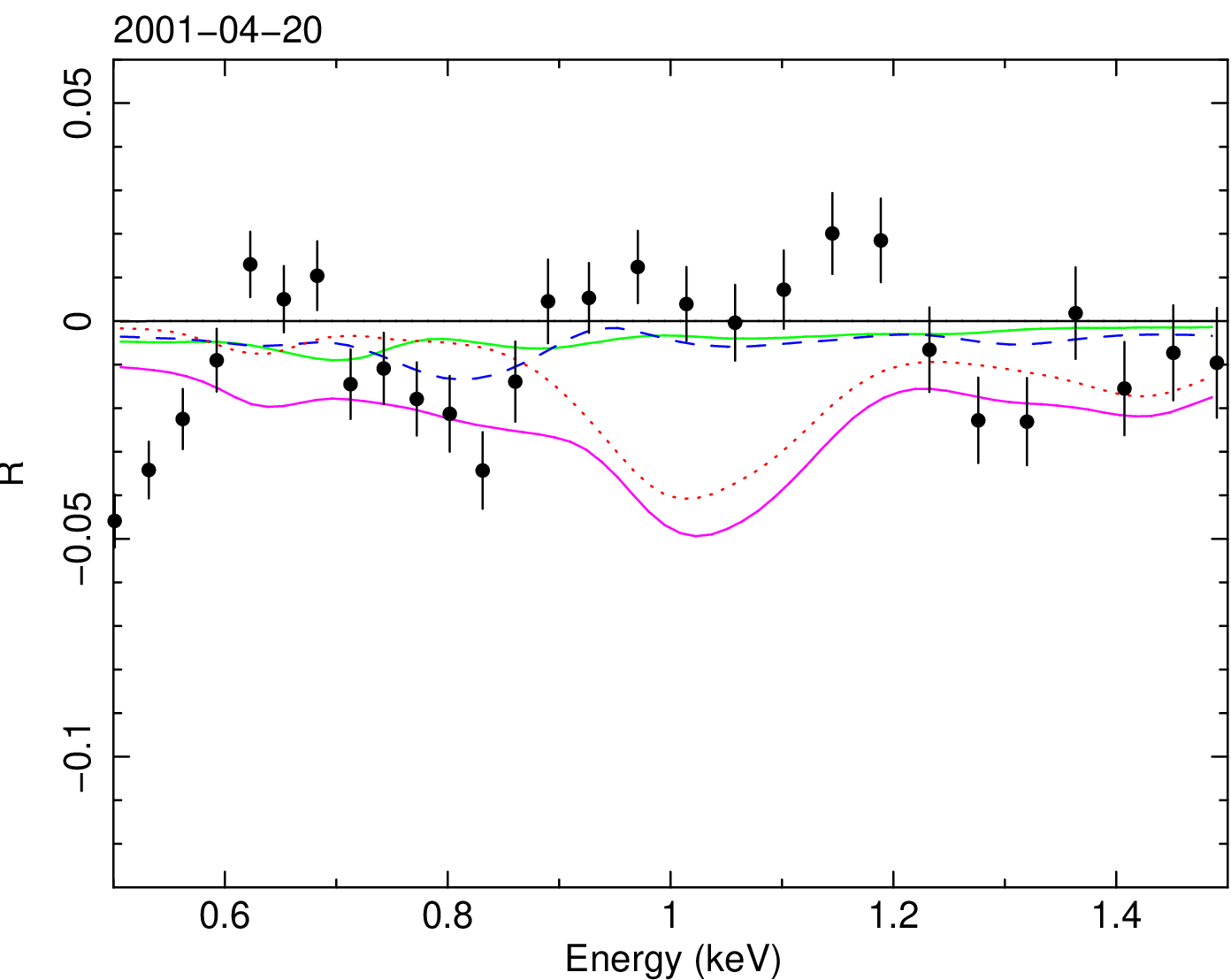}}
\caption{Data points with error bars: observed ratios of the spectra taken in
2000 (upper panel) and 2001 (lower panel) to the average spectrum taken in 2009
in the representation of Eq.~(\ref{eqn:robs}). We use here realistic
continuum models (see text) rather than simplified local power laws as used in
Fig.~\ref{fig:longterm} to take out the changes of the continuum. Solid green
lines: predicted ratio in the representation of Eq.~(\ref{eqn:rpred}) for 
ionisation component C; dashed blue lines: same for component D; dotted red
lines: same for component E; thick magenta line: components A--E combined. The
predictions for components A and B are close to zero and not shown here
separately.}
\label{fig:var11en12}
\end{figure}

According to this continuum modelling, the ionising luminosity during the 2000
and 2001 observations is lower by $-0.45$ and $-0.23$~dex. We calculated
the corresponding difference in transmission using the method described in
Sect.~\ref{sect:denslim} for the 2009 spectra. We show the results as a
comparison of the observed ratio $R_{\rm{obs}}$ to the predicted ratio
$R_{\rm{pred}}$ (cf. Eq.~\ref{eqn:robs}--\ref{eqn:rpred}) in
Fig.~\ref{fig:var11en12}.

The broad dips that are visible in the 2000 and 2001 spectral ratios shown in 
Fig.~\ref{fig:longterm} between 0.7--1.2~keV have almost disappeared. This is
due to the more realistic broad-band optical to X-ray modelling of the continua
in the present case, compared with the local 0.5--1.5~keV power-law
approximation used in Fig.~\ref{fig:longterm}. Furthermore, the spectra are not
sensitive enough to detect changes for component C. For component D, both
spectra show a narrow dip in the data around 0.8~keV. When fitted with a
Gaussian with centroid and width frozen to the values for component D taken from
Table~\ref{tab:gausblend}, we find best-fit amplitudes $N$ of $-0.014\pm0.006$
and $-0.020\pm 0.004$ for the 2000 and 2001 spectra, respectively. These values
correspond to a nominal decrease in $\xi$ of $-0.23\pm 0.10$ and $-0.33\pm
0.07$~dex. Finally, it is evident that the large changes that are predicted to
occur for component E are not present. Thus, component $E$ appears to be stable
in ionisation parameter over ten years. This can only be explained by a low
density, such that either there are no changes at all over ten years, or that
the average ionising fluxes over the few months before each of the 2000, 2001
and 2009 observations were similar by coincidence.

Comparison of the 2000 and 2001 spectra shows that the quality of the data does
not allow us to establish whether there are differences in the transmission
between these spectra.

\section{Discussion and conclusions\label{sect:discussion}}

In this paper we have constrained the distance of the dominant X-ray outflow
components in Mrk~509. This has been a challenging task, given the complexity of
the problem and the refined analysis that is necessary.

The key to achieving these results was the deep 600~ks time-averaged RGS
spectrum that allowed us to derive the ionisation structure of the outflow as
seen through X-ray absorption lines \citep{detmers2011}. The discrete components
found in that work allowed us to make predictions for the time-dependent
transmission of the outflow for different distances to the central source, and
the distance limits were obtained by comparing those predictions with the
observed spectra.

Another necessary ingredient in our analysis was the accurate broad-band
spectral modelling from the optical to the hard X-ray band
\citep{mehdipour2011}. This was not only important for deriving the shape of the
ionising continuum, but also for obtaining the underlying continuum spectrum for
each observation in the energy band around 1~keV where most of the changes in
transmission from the outflow are expected to occur. 

We have used here the continuum models as derived by \citet{mehdipour2011}.
These were based on the measured OM and pn spectra, but also on the scaled,
non-simultaneous FUSE spectra described in that paper. This gives an excellent
empirical description of the spectra (see also below). Relieving the constraints
from FUSE yields somewhat different parameters for the continuum components from
the ones used here (see Petrucci et al. 2012, in preparation), but their effects
on the present results are minor. 

Our analysis has shown that in the pn-band the continuum model spectra are
accurate down to the 1\% level (Sect.~\ref{sect:transpn}) and that narrow-band
time variations can be detected with the pn down to the 0.5\% level 
(Sect.~\ref{sect:epicpn}). For our Mrk~509 data, an instantaneous response to
the ionising continuum would predict changes up to 2\% at certain energies for
some of the ionisation components (for example between spectra 1 and 4,
Fig.~\ref{fig:delta}). Given the detection limits that we mentioned, this is
certainly a detectable signal. For the somewhat lower ionised components with
the dominant absorption troughs below $\sim$1~keV, RGS gives useful constraints
through the same blends as visible with pn, but also through some strong
individual absorption lines (\ion{Fe}{xvii}, \ion{O}{viii}).

However, our data show no evidence of an instantaneous response to the continuum
variations, as outlined in Sects.~\ref{sect:transpn} and \ref{sect:transrgs}.
Therefore we exclude the high-density limit for the gas, implying that the
absorber must be located farther away from the central black hole. Our
monitoring with Swift before the onset of the actual XMM-Newton monitoring
\citep{kaastra2011a,mehdipour2011} has been very important. It allowed us to
develop and use a realistic time-dependent photoionisation model that predicts
the change in transmission for a broad range of recombination time scales, hence
distances to the ionising source.

In our time-dependent calculations, we assumed equilibrium for the first Swift
observation as an initial condition. Because the first XMM-Newton observation
starts six weeks later, most uncertainties due to this initial condition will be
washed out for time scales shorter than this. On longer time scales this is not
the case. We note, however, that the flux during our XMM-Newton observations was
slighty higher than the long-term average flux, while the first Swift
observations was perhaps closer to this long-term average.

Again comparing our predictions with the data, we have put tight lower limits to
the distance and density of components C--E. Comparing our spectra taken in 2009
with archival data, in particular those taken in a relatively low flux state in
2000 and 2001, showed changes for component D on time scales of 8.5 years. This
is 50 times the duration of our monitoring campaign, so it yields a 50 times
lower limit to the density compared to the number given in
Table~\ref{tab:distance}, of $2.1\times 10^{8}$~m$^{-3}$, with a corresponding
upper limit to the distance of 33~pc.

In the long term, we see no detectable changes for components C and E. This does
\textsl{not} imply automatically that no changes have occurred for those
components, but we lack adequate monitoring before the 2000 and 2001
observations. Given the lack of short-term variability for components C--E, the
ionisation equilibrium depends on the time-averaged spectrum for periods of
months to years.

Because Mrk~509 can show ionising flux variations up to about 40\%, this means
in practice that for any spectrum taken, the ionising flux used to calculate the
ionisation balance for the analysis of the absorber is also uncertain by that
order of magnitude. In our case, the average flux during our XMM-Newton campaign
may have been somewhat enhanced, perhaps by 15--20\%, compared to the long-term
average, due to the outburst that peaked around the epoch of spectra 4 and 5.
This would, in retrospect, justify some minor corrections to the precise values
of $\xi$ that we derived from the time-average RGS spectrum \citep{detmers2011},
but they do not alter the general picture.

It is possible to derive upper limits to the distance using a rather simple
argument. Our lower limits to the distance $r$ imply an upper limit to the
density $n$ through $n=L/\xi r^2$. Because the typical thickness of the
absorbing layer $a$ is related to the measured column density $N_{\rm H}=na$, we
derive a lower limit to the relative thickness $a/r$. Reversing the argument,
because $a/r$ must be smaller than unity, we get an upper limit to the distance.
For our components C--E these are 9~kpc, 1.3~kpc, and 21~pc, respectively. The
limit for component C is not very useful, and for component D we have already
found an upper limit of 33~pc from the long-term variability, but the limit for
component E is interesting.

There is a caveat here, however. In this paper, we have used the five ionisation
components A--E derived from a direct fit to the total measured column
densities. \citet{detmers2011} also fitted the RGS spectrum directly using the
\textsl{xabs} model of SPEX. The components found that way are, within their
statistical uncertainties, compatible with what we use here, but for component E
there are larger uncertainties. We have used $\log\xi=3.60\pm 0.27$, while the
direct fit with the \textsl{xabs} model yielded $\log\xi = 3.26\pm 0.06$. As
discussed by \citet{detmers2011} this is the same component; the differences
arise mainly because component E is dominated in the RGS band by \ion{Fe}{xxi}
and \ion{Fe}{xxii}. Making the ionisation parameter larger but also increasing
the total hydrogen column yields almost the same ionic column densities for
these ions.

We re-ran all our procedures but with the ionisation parameters and column
densities replaced by those based on a direct \textsl{xabs} fit \citep[Table~5
in][]{detmers2011}. In particular for component E the upper limit to the
distance from the constraint $a/r<1$ increases significantly, from 21~pc to
400~pc. For the other components the differences are not very important.

Using the Chandra LETGS data of our campaign, \citet{ebrero2011} also derived
upper limits of 300 and 50~pc for the components C and the blend of D and E,
respectively. These limits are consistent with our present findings, but
\citet{ebrero2011} had to make the additional assumptions that the solid angle
of the outflow is $\sim\pi$~sr, and that the momentum of the outflow is close to
the momentum of the scattered and absorbed radiation, cf. \citet{blustin2005}.
Our present estimates are more direct. 

\citet{detmers2010} analysed the XMM-Newton spectra of Mrk~509 taken in 2005 and
2006. A simultaneous fit to the pn and RGS data of these spectra showed three
ionisation components that correspond to our components B, C, and a blend of D
and E. Components B and C did not show any significant variability between the
observations. The fits of \citet{detmers2010} indicate that the ionisation
parameter for component D/E had decreased from 3.29$\pm$0.04 in the first
observation of 2005 to 3.01$\pm$0.06 in the 2006 observation. From this change
within 0.5 year an upper limit to the distance of 0.5~pc was derived. Such a
value clearly contradicts our lower limit of about 5~pc for components D and E.
With our present model, however, a change in 0.5 year would correspond to a
distance of about 15~pc, much larger than 0.5~pc. The discrepancy can be traced
down to the use of \ion{Fe}{xii} to estimate the recombination time scale by
\citet{detmers2010}. This ion has its peak concentration near the ionisation
parameter for component D/E. It is therefore relatively insensitive to changes
in the ionising flux. As a consequence, the formal recombination time scale
\citep[Eq.~4 in][]{detmers2010} becomes very large, leading to high density and
short distance. Furthermore, the apparent change in transmission between 2005
and 2006 is possibly affected by some systematic effects and less significant
than the formal $3.9\sigma$ significance quoted above. Factors contributing to a
larger uncertainty are the lack of multi-pointing mode data for the RGS spectra
in 2005 and 2006, the shorter exposure times leading to the detection of only
three out of the five ionisation components, blending of ionisation components D
and E, and global fitting with imperfect effective area instead of using ratios
like in the present paper.  The combined difference signal from components D and
E between the first 2005 and the 2006 spectrum taken from
Table~\ref{tab:longterm} are, however, still consistent with a change at the
2.5$\sigma$ significance level. We keep it open here whether this change is real
or not, but if it is real, the upper limit of $\sim$15~pc, according to our
improved calculations, is close to the limit of 21--400~pc obtained from the
requirement that $a/r<1$ for component E as derived earlier in this section.

The distance limits derived in this paper (Table~\ref{tab:distance}) imply an
origin in the NLR or torus region of Mrk~509, rather than in the accretion
disc.  The limits are fully compatible with the upper limit of 250~pc derived
from UV variability of the fastest UV outflow component \citep{kriss2011}. This
does not exclude an accretion-disc wind component. Perhaps the highly ionised,
fast \citep[0.05$c$,][]{ponti2009}, or ultra-fast
\citep[0.14--0.20$c$,][]{cappi2009,tombesi2010} outflows found in archival
high-energy spectra of Mrk~509, but not found in our present data, could be
examples of such a disc outflow. We defer a comprehensive overview of the
various absorption components, combining both the ultraviolet and X-ray data, to
another paper (Ebrero et al. 2011, in preparation).

In summary, we derived distance limits for the main X-ray absorption components
in Mrk~509. Component C has a distance of $>70$~pc, component D is between
5--33~pc, and component E has a distance $>5$~pc but less than 21--400~pc,
depending upon modelling details. 

\begin{acknowledgements}

This work is based on observations obtained with XMM-Newton, an ESA science
mission with instruments and contributions directly funded by ESA Member States
and the USA (NASA). It is also based on observations with INTEGRAL, an ESA
project with instrument and science data centre funded by ESA member states
(especially the PI countries: Denmark, France, Germany, Italy, Switzerland,
Spain), Czech Republic, and Poland and with the participation of Russia and the
USA. This work made use of data supplied by the UK Swift Science Data Centre at
the University if Leicester. SRON is supported financially by NWO, the
Netherlands Organization for Scientific Research. J.S. Kaastra thanks the PI of
Swift, Neil Gehrels, for approving the TOO observations. M. Mehdipour
acknowledges the support of a PhD studentship awarded by the UK Science \&
Technology Facilities Council (STFC). N. Arav and G. Kriss gratefully
acknowledge support from NASA/XMM-Newton Guest Investigator grant NNX09AR01G.
Support for HST Program number 12022 was provided by NASA through grants from
the Space Telescope Science Institute, which is operated by the Association of
Universities for Research in Astronomy, Inc., under NASA contract NAS5-26555. E.
Behar was supported by a grant from the ISF. S. Bianchi, M. Cappi, and G. Ponti
acknowledge financial support from contract ASI-INAF n. I/088/06/0. P.-O.
Petrucci acknowledges financial support from CNES and the French GDR PCHE. G.
Ponti acknowledges support via an EU Marie Curie Intra-European Fellowship under
contract no. FP7-PEOPLE-2009-IEF-254279. K. Steenbrugge acknowledges the support
of Comit\'e Mixto ESO - Gobierno de Chile.

\end{acknowledgements}

\bibliographystyle{aa}
\bibliography{paper}

\end{document}